\documentclass{article} 
\usepackage[final]{colm2025_conference}

\usepackage{microtype}
\usepackage{hyperref}
\usepackage{url}
\usepackage{booktabs}
\usepackage{amsmath, amsfonts, amsthm, bm}
\usepackage{soul}
\usepackage{tabularx, multirow}
\usepackage{dsfont}
\usepackage[ruled,vlined,linesnumbered]{algorithm2e}
\usepackage{xcolor, colortbl}
\usepackage{xspace}
\usepackage{resizegather}
\usepackage{makecell}
\usepackage[normalem]{ulem}
\usepackage{CJKutf8}
\usepackage{enumitem}
\usepackage{arydshln}
\usepackage{enumitem}
\usepackage{balance}
\usepackage{svg}
\usepackage{tikz}
\usepackage{pifont}
\usepackage{wasysym}
\usepackage{mathtools}
\usepackage{lipsum}
\usepackage{graphicx}
\usepackage{booktabs}
\usepackage{caption, subcaption}
\usepackage{lineno}
\usepackage{graphicx}
\usepackage{wrapfig}
\usepackage{lipsum} 
\definecolor{darkblue}{rgb}{0, 0, 0.5}
\hypersetup{colorlinks=true, citecolor=darkblue, linkcolor=darkblue, urlcolor=darkblue}

\title{Imagine All The Relevance: Scenario-Profiled Indexing \\ with Knowledge Expansion for Dense Retrieval}

\newcommand{\aspace}{\hspace{1em}}
\newcommand{\ys}{$^{\heartsuit}$}
\newcommand{\ko}{$^{\clubsuit}$}

\author{%
    Sangam Lee\ys\aspace
    Ryang Heo\ys\aspace
    SeongKu Kang\ko\aspace
    Dongha Lee\ys\aspace\vspace{3pt}\\
    \ys Yonsei University\aspace\ko Korea University \\ 
    \texttt{\{salee, ryang1119,donalee\}@yonsei.ac.kr}\aspace\texttt{seongkukang@korea.ac.kr}
}

\usepackage{titletoc}
\newcommand\DoToC{
  \startcontents
  \printcontents{}{1}{\textbf{Contents of Appendix}\vskip3pt\hrule\vskip5pt}
  \vskip3pt\hrule\vskip5pt
}

\usepackage{xspace}
\newcommand{\spike}{\textsc{SPIKE}\xspace}

\begin{document}

\ifcolmsubmission
\linenumbers
\fi

\maketitle

\begin{abstract}
Existing dense retrieval models struggle with reasoning-intensive retrieval task as they fail to capture implicit relevance that requires reasoning beyond surface-level semantic information.
To address these challenges, we propose Scenario-Profiled Indexing with Knowledge Expansion (\spike), a dense retrieval framework that explicitly indexes implicit relevance by decomposing documents into scenario-based retrieval units. 
\spike organizes documents into scenario, which encapsulates the reasoning process necessary to uncover implicit relationships between hypothetical information needs and document content.
\spike constructs a scenario-augmented dataset using a powerful teacher large language model (LLM), then distills these reasoning capabilities into a smaller, efficient scenario generator. 
During inference, SPIKE incorporates scenario-level relevance alongside document-level relevance, enabling reasoning-aware retrieval. 
Extensive experiments demonstrate that \spike consistently enhances retrieval performance across various query types and dense retrievers. 
It also enhances the retrieval experience for users through scenario and offers valuable contextual information for LLMs in retrieval-augmented generation (RAG).
\end{abstract}

\section{Introduction}
\label{sec:intro}
Information retrieval (IR) systems are essential for helping users find relevant information within the overwhelming volume of available data. 
Over the years, dense retrieval~\citep{EMNLP/KarpukhinOMLWEC20/DPR, sigir/KhattabZ20/ColBERT} has emerged as a dominant approach. 
It employs pre-trained language models (PLMs) to encode queries and documents into shared vector spaces, enabling a deeper understanding of their semantic relationships. 
Despite these advancements, there remain fundamental challenges in IR.

\begin{figure}[!t]
    \centering
    \includegraphics[width=\columnwidth]{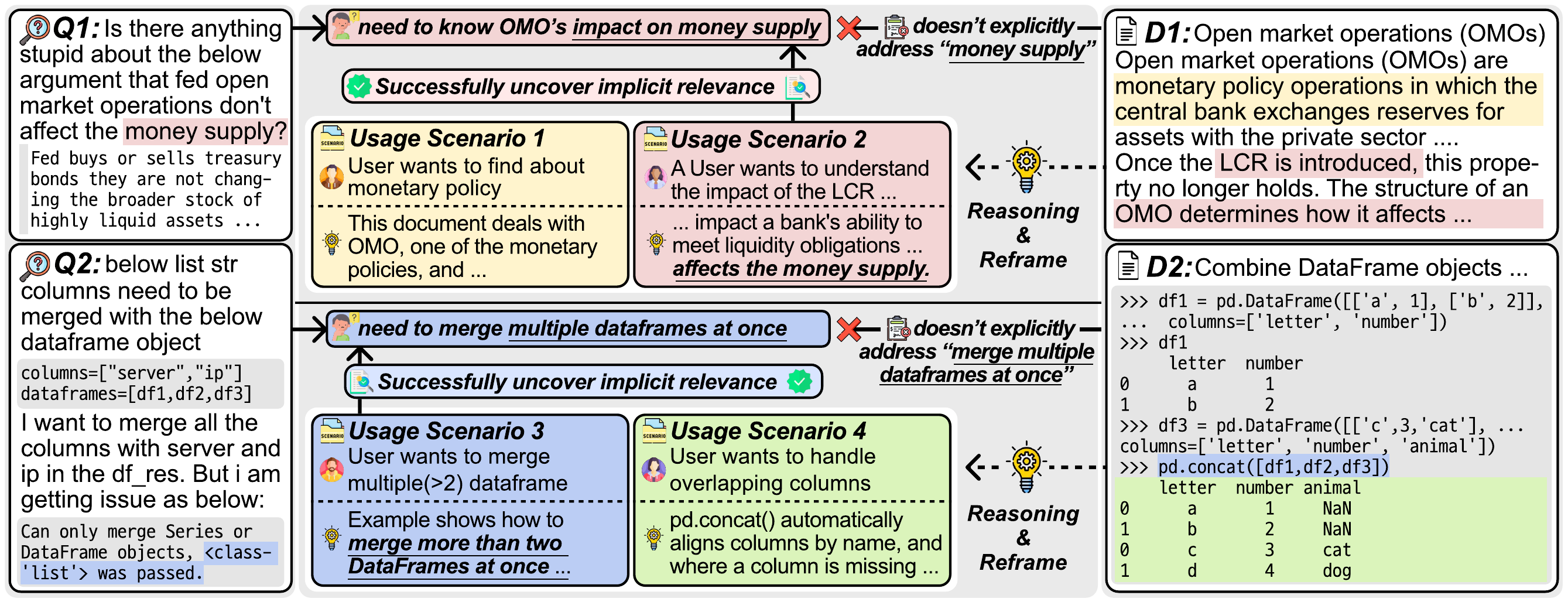}
    \caption{Existing retrieval methods fail to capture implicit relevance which requires intensive reasoning, as they encode document into single vector without any reasoning. In contrast, \spike introduces scenario, explicitly modeling how a document establishes relevance to potential information needs.}
    \label{fig:motivation}
\end{figure}

One of the most significant challenges is that \textbf{dense retrieval struggles to capture deeper implicit relevance beyond surface-level semantic.}
Recently, BRIGHT~\citep{Su2024BRIGHTAR}, a benchmark for reasoning-intensive retrieval tasks, has been proposed.
Unlike traditional IR benchmarks such as BEIR~\citep{Thakur2021BEIRAH} and MTEB~\citep{Muennighoff2022MTEBMT}, it requires intensive reasoning to uncover implicit relevance between a query and relevant documents, which cannot be captured through simple keyword or surface-level semantic information.
For example, in Figure~\ref{fig:motivation} (Upper), \textit{Q1} asks about the impact on Open Market Operations (OMOs) on money supply, whereas its relevant document \textit{D1} doesn't explicitly address about it. 
Instead, it discusses OMOs' influence on the Liquidity Coverage Ratio (LCR), which in turn affects money supply.
To uncover implicit relevance (OMO → LCR → Money Supply), it is \textbf{necessary to reason from the LCR-related information} of \textit{D1}, which discusses regulatory policies and financial mechanisms. 
However, existing dense retrievers lack the capability to perform reasoning, and thus fail to uncover such implicit relationships.
As a result, they struggle with reasoning-intensive retrieval task.~\citep{Su2024BRIGHTAR}.

\textbf{This limitation becomes even more pronounced when they handle query-document pairs of significantly different formats}, such as code and natural language. 
As shown in Figure~\ref{fig:motivation} (Lower), \textit{D2} provides the necessary information for \textit{Q2} only in the form of code examples (e.g., ``pd.concat([df1, df2, df3])''). 
In such cases, bridging the semantic gap between the natural language query and the code-based document requires reasoning over specific parts of the code to uncover implicit relevance. 
However, existing dense retrieval models are unable to effectively address this discrepancy because they are primarily trained on natural language query-document pairs~\citep{reimers2019sentence, bge_embedding, meng2024sfrembedding}, making them ill-suited for bridging the gap between them. 

In this work, we propose \textbf{S}cenario-\textbf{P}rofiled \textbf{I}ndexing with \textbf{K}nowledge \textbf{E}xpansion (\textbf{\spike}), a dense retrieval framework that explicitly indexes potential implicit relevance within documents.
The key idea of \spike is to reframe document representations into \textbf{hypothetical retrieval scenarios}, where each scenario encapsulates the reasoning process required to uncover implicit relevance between a hypothetical information need and the document content. 
As illustrated in Figure~\ref{fig:motivation}, \spike organizes document knowledge into hypothetical retrieval scenarios, which are considered alongside the document during the retrieval process.
This approach 1) enhances retrieval performance by explicitly modeling how a document addresses hypothetical information needs, capturing implicit relevance between query and document. 
It also 2) effectively connects query-document pairs across different formats such as code snippets, enabling semantic alignment despite format differences.
Additionally, it 3) enhances the retrieval experience for users by providing useful information while also serving as valuable context for LLMs in RAG settings.

Specifically, \spike consists of the following three steps: 1) constructing a scenario-augmented training dataset using a teacher LLM to generate high-quality supervision, 2) employing scenario distillation to transfer the reasoning capabilities of teacher LLM into a smaller, more efficient scenario generator, and 3) using the trained scenario generator to formulate structured retrieval scenarios for documents.
During inference, \spike considers scenario-level relevance alongside document-level relevance to retrieve the relevant documents.
Our extensive experiments demonstrate that \spike not only improves retrieval performance but also enhances the retrieval experience for users. 
Additionally, we demonstrate that \spike serves as a valuable additional context for LLMs in RAG settings.
For reproducibility, our codes are publicly available at the anonymous github repository.\footnote{\url{https://github.com/augustinLib/SPIKE}}

The main contributions of our work are summarized as follows:
\begin{itemize}[leftmargin=*,topsep=2pt,itemsep=2pt,parsep=0pt]
  \item We propose \spike, a dense retrieval framework that decomposes documents into scenarios. These scenarios enable effective retrieval by capturing implicit relevance.
  \item Our extensive experiments show that \spike consistently improves performance across diverse retrieval models, query types and document types.
  \item \spike helps users by providing explanations that make retrieved results easier to understand, while also making it easier for LLMs to generate accurate answers in RAG.
\end{itemize}

\section{Related Works}
\label{sec:relatedworks}
\textbf{Reasoning-intensive retrieval.}
Traditional retrieval benchmarks~\citep{Thakur2021BEIRAH, Muennighoff2022MTEBMT} have largely focused on surface-level information-seeking queries where simple keyword or semantic matching-based retrieval is often sufficient.
To address this limitation of traditional retrieval benchmarks, \cite{Su2024BRIGHTAR} propose BRIGHT, a benchmark that requires reasoning to retrieve relevant documents for a query. 
\cite{Su2024BRIGHTAR} and \cite{Niu2024JudgeRankLL} propose LLM-based query expansion and reranking as potential solutions for reasoning-intensive retrieval task.
However, they have several limitations. 
First, LLM-based query expansion and reranking introduce significant computational overhead. 
Since these approaches require running inference on inefficient LLMs ($>$ 8B parameters) for every query, they are computationally expensive and lead to high latency. 
Second, rerankers are dependent on the first-stage retrieval performance. 
If the first-stage retrieval fails to retrieve relevant documents, even a strong reranker cannot recover them. 
This dependency prevents rerankers from fully addressing the challenges of reasoning-intensive retrieval.
Overall, these limitations highlight the importance of improving first-stage retrieval for reasoning-intensive retrieval tasks, though this area remains largely underexplored.

\textbf{Document expansion \& organization.}
~Prior works have improved retrieval performance by appending pseudo queries~\citep{Nogueira2019DocumentEB, Chen2024ReInvokeTI}, summaries~\citep{Jeong2021UnsupervisedDE}, or keyphrases~\citep{Boudin2020KeyphraseGF} to the original document.
Since these approaches are performed at the indexing stage, they do not introduce additional inference-time overhead.
Another line of works replace the original document representations with more effective retrieval units, such as summaries~\citep{Sarthi2024RAPTORRA} or propositions~\citep{Chen2023DenseXR}.
While these approaches refine document representations, they are limited in handling implicit information that cannot be addressed through simple semantic matching.
As a result, they struggle in reasoning-intensive retrieval task.
Additionally, the models used in these methods are typically trained only on natural language documents, they cannot be directly applied to non-natural language documents like code snippet.

\section{Proposed Method: \spike}
\label{sec:method}
In this section, we present a dense retrieval framework, \textbf{S}cenario-\textbf{P}rofiled \textbf{I}ndexing with \textbf{K}nowledge \textbf{E}xpansion (\spike), which introduces a scenario-profiled retrieval to explicitly index potential implicit relevance.
The overall framework is illustrated in Figure~\ref{fig:method}.

\subsection{Scenario: Reasoning format for modeling implicit relevance}
\label{subsec:documentscenario}
The first step of our \spike framework is to define the concept of a scenario, which serves as a structured reasoning format used to explicitly model the implicit relevance between a document and potential information needs.
\spike reframes each document into multiple scenarios, each representing a distinct reasoning path that explains how the document could satisfy a hypothetical information need.
Through scenario generation, \spike uncovers the diverse forms of implicit relevance a document may hold.
To generate meaningful scenarios, we employ an LLM-driven reasoning approach that analyzes the document and constructs each scenario through a step-by-step process involving the following scenario components:

\begin{figure}[!t]
    \centering
    \includegraphics[width=\columnwidth]{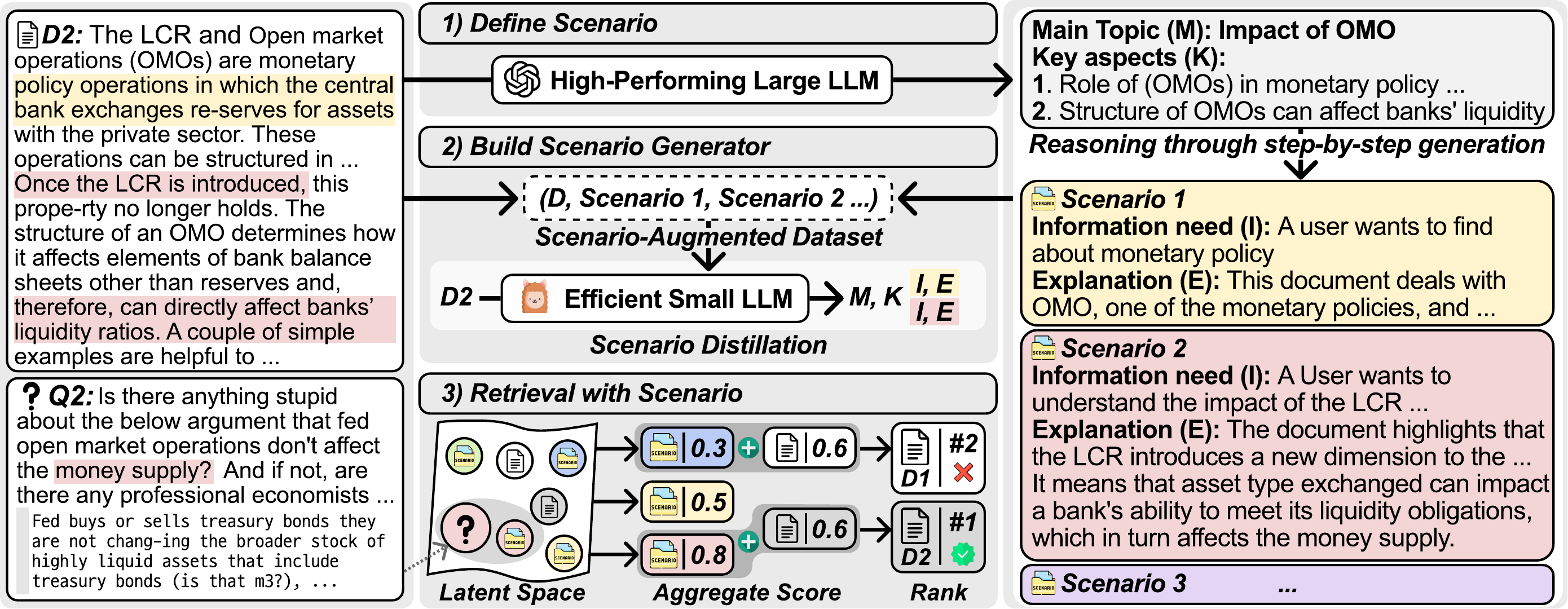}
    \caption{Overview of \spike framework. (1) \spike define Scenario and generate it with high-performing large LLM. (2) Then, it construct scenario-augmented training set, and use this to optimize the efficient student LLM. During inference, (3) \spike considers scenario-level relevance alongside document-level relevance to retrieve the documents.}
    \label{fig:method}
\end{figure}

\textbf{Main topic (M).}
~To construct meaningful scenarios, we first identify the main topic of the document, which serves as a high-level summary of its content. 
This ensures that subsequent scenario components remain grounded in the document’s overall theme, preventing them from diverging too far from its core subject. 

\textbf{Key aspects (K).}
~Then, we extract key aspects that capture the diverse multi-aspects of the document’s content. 
Key aspects provide a more detailed breakdown of the document’s content compared to the main topic, capturing the diverse and specific information embedded within the document. 
By explicitly listing various details at this step, subsequent steps can produce a broader range of information, ensuring diverse scenario coverage.

\textbf{Information needs (I).}
~The next step is to generate information needs that reflect the potential retrieval intents a document can address.
Specifically, we generate diverse information needs that describe situations in which the given document can effectively address hypothetical information requests based on the main topic and each key aspects.

\textbf{Explanations (E).}
~For each generated information need, we generate an explanation that explicitly convey the connection between the document and the generated information need. 
These explanations serve as the core component for modeling relevance, as they describe why and how a document provides the necessary information to satisfy a given information need.
This step ensures that each scenario captures implicit relevance by explicitly linking the document’s content to the generated information need. 

By generating components autoregressively, \spike uncovers potential implicit relevance within a document, ultimately modeling it through \textit{Explanations (E)}.
Leveraging \textit{E} for retrieval enables \spike to overcome limitations of existing dense retrieval models, effectively capturing implicit relevance through explicit reasoning.
However, indexing only \textit{E} may lead to dense vector representations that fail to fully capture the document’s overall context.
To mitigate this issue, \spike incorporates the \textit{Main topic (M)} component along with \textit{E}, indexing their combination \textit{(M+E)}.
The main topic provides a high-level anchor that maintains coherence with the document's overall content, while explanations explicitly model the reasoning processes essential for identifying implicit relevance.
This balanced approach ensures that retrieval representations remain both contextually grounded and reasoning-aware, ultimately enhancing retrieval performance more effectively.

\subsection{Scenario generator \& Scenario Indexing}
\label{subsec:buildscenariogenerator}
Since scenario generation requires strong reasoning capabilities, high-performing LLMs like GPT-4o are essential for producing high-quality scenarios. 
However, applying such models to an entire corpus is computationally expensive and impractical for large-scale corpus. 
While smaller open-source models are more efficient, they often lack the reasoning ability needed for scenario generation. 
To address this, we first (1) use a high-performing LLM to construct a scenario-augmented training dataset with high-quality supervision, then (2) employ scenario distillation to train a smaller model, effectively transferring the reasoning capabilities of large LLMs into a small scenario generator.

\textbf{Scenario-augmented training dataset.}
~To train an effective scenario generator, we first construct a scenario-augmented training dataset $\mathcal{D} = \{(d, \tilde{S}^d)_i\}_{i=1}^{N}$, where each document $d$ is paired with a sequence of scenarios $\tilde{S}^{d} = \{\tilde{s}_1, \dots \tilde{s}_k\}$.
Since generating high-quality scenarios demands reasoning that goes beyond basic text generation, we leverage high-performing LLM such as GPT-4o to construct a scenario-augmented training dataset. 

\textbf{Scenario distillation.}
~Once the scenario-augmented training dataset is constructed, we train a smaller scenario generator to efficiently produce reasoning-driven scenarios. 
Specifically, we minimize the following distillation loss: 
\begin{equation}
\label{eq:distilllossterm}
    \mathcal{L}_{\text{Distillation}} = -\sum \log P(\tilde{S}^{d} \mid \mathcal{I}, d; \theta)
\end{equation}
where $\theta$ denotes the parameters of the scenario generator model, trained to generate scenarios $\tilde{S}^{d}$ given the document $d$ and instruction for scenario generation $\mathcal{I}$.

\textbf{Scenario indexing.}
~After training the scenario generator, we generate a set of scenarios for each document $d$ in the corpus.
From each generated scenario, we extract only the Main topic (M) and Explanation (E) components, as discussed in Section~\ref{subsec:documentscenario}, which are combined to construct the final scenario representation set $S^d=\{s_1,\dots s_k\}$.
Then each scenario representation is encoded into a dense vector using the same encoder $\mathcal{E}$ used for document representations. 
These vectors are used to build a scenario-profiled index alongside the standard document index. 
This additional scenario-profiled index allows the retrieval system to capture implicit relevance more effectively by leveraging reasoning-derived scenario representations during retrieval. 
Since scenario generation and indexing are performed offline, this approach imposes no additional inference-time burden, unlike query expansion~\citep{Su2024BRIGHTAR} or LLM-based reranking methods~\citep{Niu2024JudgeRankLL}.

\subsection{Retrieval with scenario}
\label{subsec:inference}
During retrieval, we produce the final ranked list $Y_{\text{final}}$. 
To this end, we first compute the relevance scores between the query and each document, as well as its associated scenarios with pre-built indexes.
Let $q$ be the query and $\mathcal{E}$ the dense retrieval model. 
For a given document $d$ with associated scenario set $S^d$, we compute relevance scores as follow:
\begin{equation}
\begin{aligned}
r_d &= \text{sim}(\mathcal{E}(q), \mathcal{E}(d)), \quad
r_{s} = \text{sim}(\mathcal{E}(q), \mathcal{E}(s)), \quad \forall\, s \in S^d
\end{aligned}
\label{eq:relevance_grouped}
\end{equation}
where $\text{sim}(\cdot)$ denotes cosine similarity, $r_d$ represents the relevance score for the document, and $r_{s}$ represents the relevance score for the scenario.
Then, we select the maximum relevance score among the scenarios associated with each document and compute the final relevance score as a weighted sum of the document and scenario scores:
\begin{equation}
\begin{aligned}
r_{\text{final}}(d) = \alpha\, r_d + (1-\alpha)\, \max_{s \in S^d} \{r_{s}\}
\label{eq:finalrankedlist}
\end{aligned}
\end{equation}
where $\alpha$ is the relevance weight, which is the hyperparameter that controls the effect of document and scenario. 
Finally, we produce $Y_\text{final}$ by calculating final relevance score $r_{\text{final}}(d)$ for all documents and sorting them in descending order.

\subsection{Efficient retrieval strategy}
\label{subsec:efficientretrievalstrategy}
While \spike's scenario-profiled index enhances retrieval effectiveness, a naive implementation could introduce significant latency by scoring every scenario for all documents. 
Such an approach would cause additional overhead and latency to scale proportionally with the corpus size ($N$). 
To ensure practical efficiency, we employ an efficient retrieval strategy. 
For a given query, \spike first identifies a candidate set of top-k' documents using only the document scores ($r_d$). 
Subsequently, scenario scores ($r_s$) are computed exclusively for this limited subset. 
This approach ensures that the additional computation is bounded by the hyperparameter $k'$ (where $k < k' \ll N$) and does not scale with the corpus size.

\section{Experiments}
\label{sec:experiments}
In this section, we conduct our experiments to answer the following research questions:
\begin{itemize}[leftmargin=*,topsep=2pt,itemsep=2pt,parsep=0pt]
    \item \textbf{RQ1:} Can \spike effectively enhance the retrieval performance? 
    \item \textbf{RQ2:} Can \spike's scenarios serve as useful information for real-world users?
    \item \textbf{RQ3:} Can \spike's scenarios serve as an effective additional context in a RAG setting?
\end{itemize}

\subsection{Experimental settings}
\label{subsec:experimentsetting}

\textbf{Datasets \& Evaluation metric.}
~We use BRIGHT benchmark~\citep{Su2024BRIGHTAR} to assess retrieval performance of \spike.
By evaluating on BRIGHT, we assess how much \spike improves the performance of dense retrieval models that previously struggled on reasoning-intensive retrieval task, demonstrating its effectiveness.
As done in previous works, we evaluate retrieval performance using nDCG@10.
More details are provided in Appendix~\ref{appendix:evaldataset}.

\textbf{Backbone models.}
~To demonstrate that \spike can be applied effectively across different dense retrievers, we evaluate performance of \spike with 6 representative dense retrievers.
Specifically, we conduct experiments with BGE-Large~\citep{bge_embedding}, SBERT~\citep{reimers2019sentence}, E5-Mistral-7B~\citep{Wang2023ImprovingTE}, SFR-Embedding-Mistral~\citep{meng2024sfrembedding}, GRIT~\citep{muennighoff2024GRIT} and gte-Qwen1.5~\citep{li2023towards}.

\textbf{Implementation details.}
~We construct the scenario-augmented training dataset by randomly sampling 300 documents per dataset from StackExchange split of BRIGHT~{\citep{Su2024BRIGHTAR}} and all datasets in BEIR~\citep{Thakur2021BEIRAH}, resulting in a total of 8,100 documents. 
Note that there is not any exposure to test queries or their relevance annotations, ensuring that retrieval evaluation remains entirely independent of the training process.
For each sampled document, we use GPT-4o to generate scenarios, ensuring that the dataset contains high-quality scenarios that facilitate deeper reasoning over a document’s contents. 
For the scenario generator, we use Llama-3.2-3B-Instruct~\citep{Dubey2024TheL3} as the backbone model and fine-tune it using LoRA~\citep{Hu2021LoRALA}.
For our main experiments, we set the relevance weight in Equation~\eqref{eq:finalrankedlist} to 0.7.
We use the efficient retrieval strategy mentioned in Section~\ref{subsec:efficientretrievalstrategy} for all experiments, setting the value $K'$ at 1000.
For reproducibility, we also provide more details about implementation details in Appendix~\ref{appendix:implementationdetails}.

\begin{table}[t!]
\setlength{\tabcolsep}{4pt}
\centering

\resizebox{\textwidth}{!}{
    \begin{tabular}{lcccccccccccccc}
    \toprule
    & \multicolumn{5}{c}{\textbf{Natural language}} 
    & \multicolumn{4}{c}{\textbf{Code}} 
    & \multicolumn{3}{c}{\textbf{Math}} 
    & \multirow{2}{*}{\centering \textbf{{Avg.}}} & \multirow{2}{*}{\centering \textbf{Improv.}} \\
    \cmidrule(r){2-6} \cmidrule(r){7-10} \cmidrule(r){11-13}
    & \textbf{Bio.} & \textbf{Earth.} & \textbf{Econ.} & \textbf{Psy.} & \textbf{Sus.} 
    & \textbf{Rob.} 
    & \textbf{Stack.} & \textbf{Leet.} & \textbf{Pony} 
    & \textbf{Aops} & \textbf{TheoQ.} & \textbf{TheoT.} \\
    \midrule
    \multicolumn{15}{c}{\textit{Dense retrieval models (\(<1\)B)}} \\
    \midrule
    BGE 
    & 12.0 & 24.2 & 16.6 & 17.4 & \textbf{13.3} 
    & \textbf{12.2 }
    & 9.5 & 26.7 & 5.6 
    & \textbf{6.0} & 13.0 & 6.9 
    & 13.6 & \multirow{2}{*}{\centering \textbf{{+5.9\%}}} \\
    
    +\spike 
    & \textbf{13.2} & \textbf{26.4} & \textbf{17.0} & \textbf{18.1} & 13.2 
    & 11.5 
    & \textbf{13.3} & \textbf{27.1} & \textbf{6.4 }
    & 4.8 & \textbf{13.0} & \textbf{8.5 }
    & \textbf{14.4} & \\
    
    \cdashline{1-15}
    \addlinespace[2pt]
    
    SBERT 
    & 15.5 & 20.1 & 16.6 & \textbf{22.6} & 15.3 
    & 8.4 
    & 9.5 & 26.4 & 6.9 
    & 5.3 & \textbf{20.0} & 10.8 
    & 14.8 & \multirow{2}{*}{\centering \textbf{{+6.1\%}}} \\
    
    +\spike 
    & \textbf{18.2} & \textbf{23.1} & \textbf{17.9} & 21.3 & \textbf{15.5}
    & \textbf{9.0}
    & \textbf{13.4} & \textbf{26.7} & \textbf{8.1} 
    & \textbf{5.4} & 19.3 & \textbf{11.2} 
    & \textbf{15.7} & \\
    
    \midrule
    \multicolumn{15}{c}{\textit{Dense retrieval models (\(>1\)B)}} \\
    \midrule
    
    E5-Mistral 
    & 18.8 & 26.0 & 15.5 & 15.8 & 18.5
    & 16.4
    & 9.8 & 28.7 & 4.8
    & \textbf{7.1} & \textbf{26.1} & 26.8
    & 17.9 & \multirow{2}{*}{\centering \textbf{{+20.7\%}}} \\
    
    +\spike 
    & \textbf{25.9} & \textbf{33.0} & \textbf{18.2} & \textbf{20.6} & \textbf{20.6}
    & \textbf{18.4}
    & \textbf{16.2} & \textbf{29.4} & \textbf{17.5}
    & 7.0 & 23.4 & \textbf{28.4}
    & \textbf{21.6} & \\
    \cdashline{1-15}
    \addlinespace[2pt]
    
    SFR 
    & 19.5 & 26.6 & 17.8 & 19.0 & 19.8
    & 16.7
    & 12.7 & 27.4 & 2.0
    & \textbf{7.4} & \textbf{24.3} & 26.0
    & 18.3 & \multirow{2}{*}{\centering \textbf{{+18.6\%}}}\\
    
    +\spike 
    & \textbf{23.6} & \textbf{31.7} & \textbf{19.9} & \textbf{26.0} & \textbf{21.2}
    & \textbf{17.8}
    & \textbf{17.6} & \textbf{28.6} & \textbf{17.3}
    & 6.5 & 22.8 & \textbf{27.5}
    & \textbf{21.7} & \\
    \cdashline{1-15}
    \addlinespace[2pt]
    
    GRIT 
    & 25.0 & \textbf{32.8} & 19.0 & 19.9 & 18.0
    & 17.3
    & 11.6 & 29.8 & \textbf{22.0}
    & 8.8 & 25.1 & 21.1
    & 20.9 & \multirow{2}{*}{\centering \textbf{{+4.3\%}}}\\
    
    +\spike 
    & \textbf{27.8} & 29.0 & \textbf{20.0} & \textbf{20.4} & \textbf{19.0}
    & \textbf{19.2}
    & \textbf{16.7} & \textbf{32.0} & 18.3
    & \textbf{9.2} & \textbf{25.2} & \textbf{24.9}
    & \textbf{21.8} & \\
    \cdashline{1-15}
    \addlinespace[2pt]
    
    Qwen 
    & 30.9 & 36.2 & 17.7 & 24.6 & 14.9
    & 13.5
    & 19.9 & 25.5 & 14.4
    & \textbf{27.8} & \textbf{32.9} & \textbf{32.9}
    & 24.3 & \multirow{2}{*}{\centering \textbf{{+3.3\%}}}\\
    
    +\spike 
    & \textbf{32.4} & \textbf{41.2} & \textbf{23.7} & \textbf{25.7} & \textbf{24.7}
    & \textbf{16.0}
    & \textbf{23.7} & \textbf{26.3} & \textbf{16.7}
    & 12.5 & 27.1 & 31.0
    & \textbf{25.1} & \\
    
    \bottomrule
    \end{tabular}
}
\caption{The retrieval performance of existing retrieval models and our \spike framework on the BRIGHT benchmark with the original query. We report nDCG@10 for all datasets. Avg. denotes the average score across 12 datasets and Improv. denotes the improvement rate of the average score. The best score on each model is shown in bold.}
\label{tab:main_results}
\end{table}

\subsection{\spike improves retrieval performance (RQ1)}
\label{subsec:mainresult}

\textbf{Main result on BRIGHT.}
~Table \ref{tab:main_results} shows the retrieval performance of dense retrieval models on the BRIGHT benchmark, comparing their original performance against their \spike-enhanced versions.
Across different retrieval models and datasets, \spike consistently improves average retrieval performance, demonstrating its effectiveness in capturing implicit relevance through scenario-profiled indexing. 
These improvements are observed across both small ($<$1B) and large ($>$1B) dense retrieval models, highlighting the ability to generalize across different retrieval architectures. 
Notably, \spike yields substantial gains for models with relatively weaker baseline performance, such as E5-Mistral and SFR, where retrieval accuracy improves by over 18\%, emphasizing its potential to bridge reasoning gaps in weaker LLM-based retrieval models.
Moreover, the results show that \spike provides significant benefits in datasets involving non-natural language content, such as code-based datasets. 
Standard dense retrieval models struggle in these datasets due to the inherent discrepancy between non-natural language documents and natural language queries. 
\spike mitigates this gap and connects between them by leveraging scenario-profiled retrieval, which explicitly describes in natural language how a non-natural language document addresses diverse information needs. 
In the math domain, \spike also shows performance gain. 
In \textbf{TheoT.}, where documents feature LaTeX-formatted mathematical expressions, our scenario-profiled approach substantially improves retrieval performance by clarifying implicit reasoning steps.
In contrast, the performance gains in \textbf{Aops.} and \textbf{TheoQ.} are less pronounced, primarily because these datasets present documents in a question-like format.
Since these documents are already written in a question-like format, it becomes challenging to generate additional scenarios that meaningfully enrich them.

\begin{figure}[t]
    \centering
    \begin{minipage}{0.48\textwidth}
        \centering
        \includegraphics[width=\textwidth]{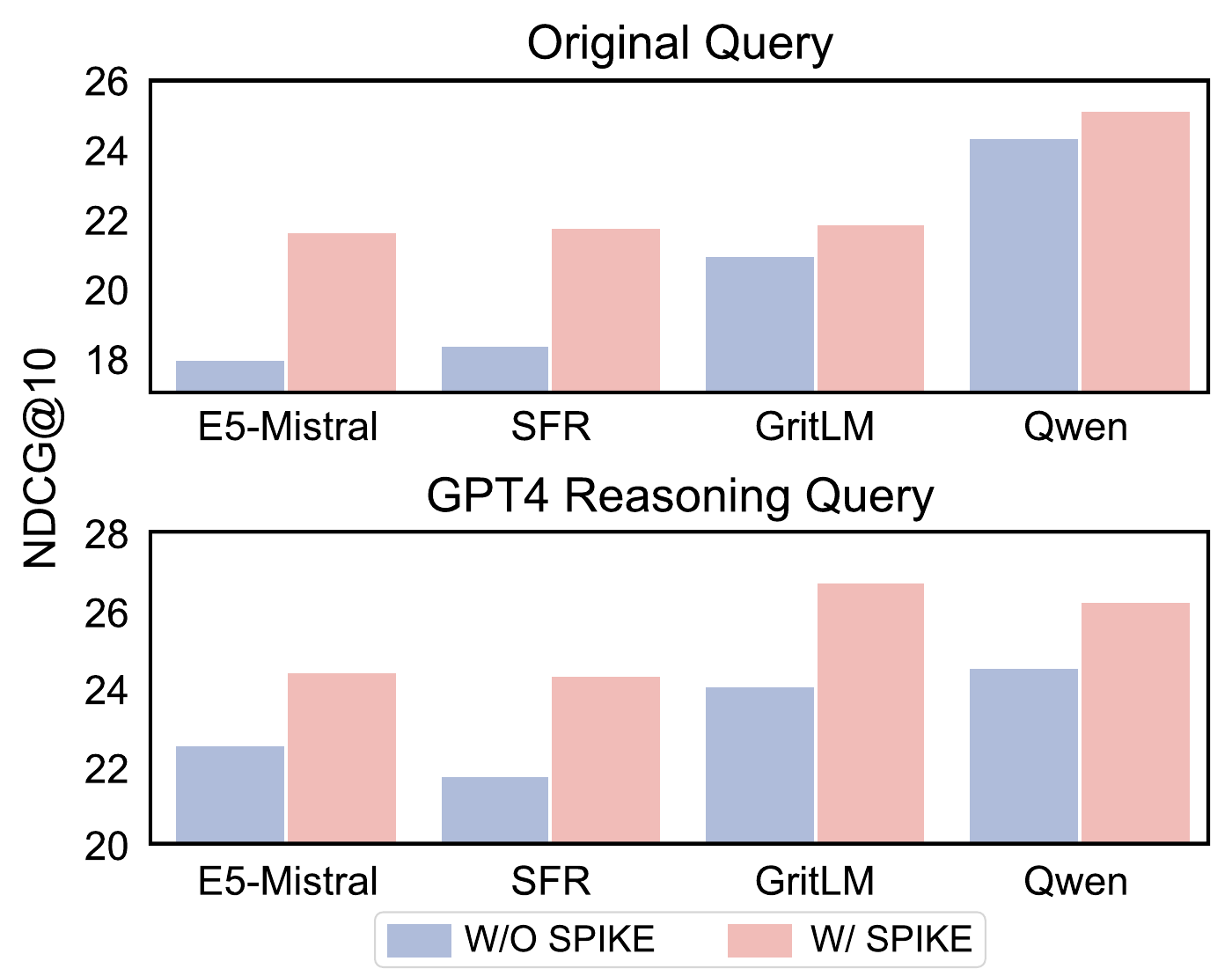}
        \caption{Average nDCG@10 scores on BRIGHT for different query types.}
        \label{fig:gpt4query}
    \end{minipage}
    \hfill
    \begin{minipage}{0.48\textwidth}
        \centering
        \includegraphics[width=\textwidth]{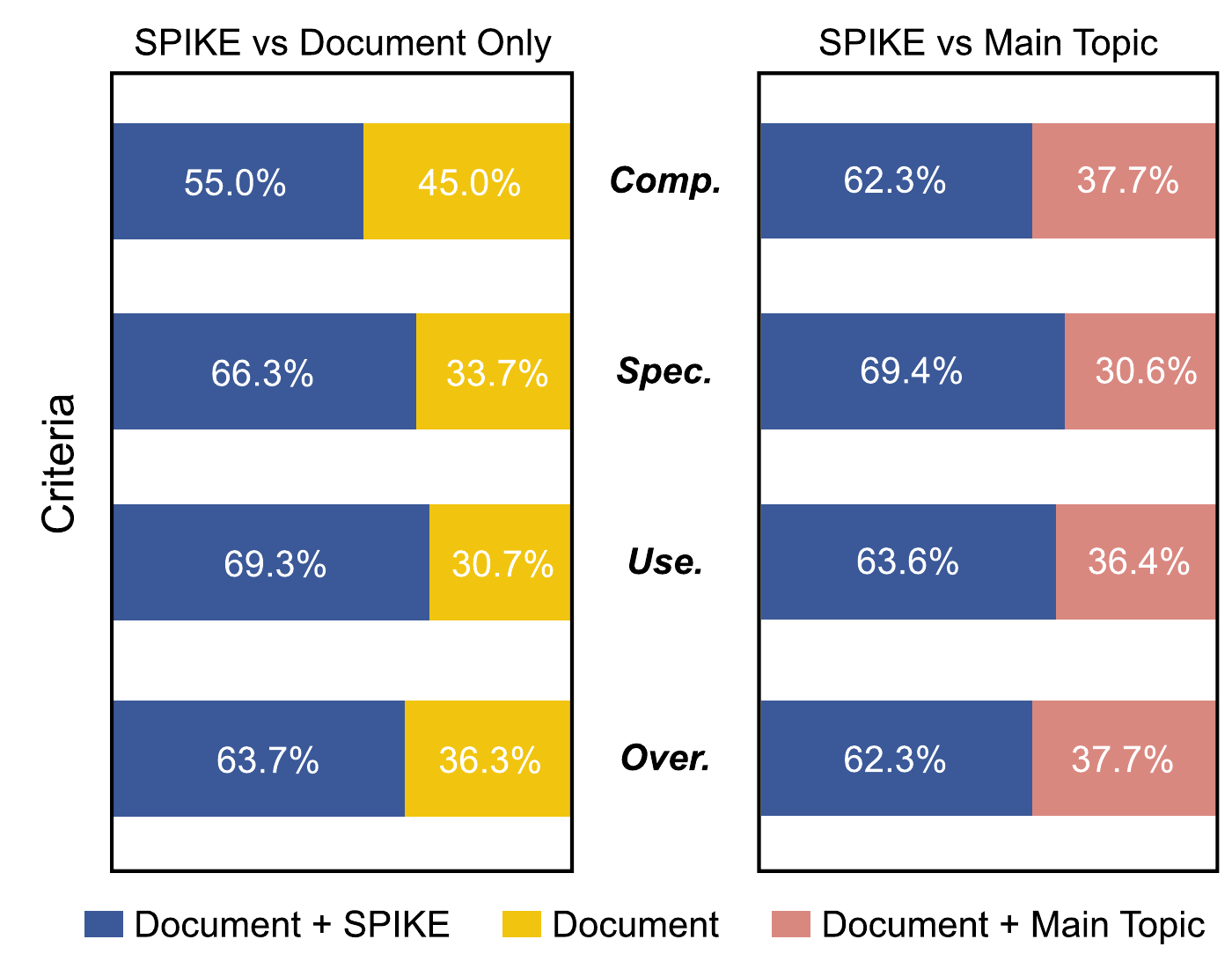}
        \caption{Human evaluation of pairwise comparisons for retrieval results.}
        \label{fig:human_eval}
    \end{minipage}
\end{figure}

\textbf{Results with reasoning-augmented queries.}
~We additionally evaluate performance using reasoning-augmented queries, obtained by prompting LLM to reformulate the original queries with explicit step-by-step reasoning~\citep{Su2024BRIGHTAR}.
This setup allows us to assess how \spike performs when queries already contain explicit reasoning steps.
In this experiment, we use GPT-4 reasoning queries provided by the BRIGHT benchmark, which are reformulated from the original queries in BRIGHT.
For further details, please refer to the Appendix~\ref{appendix:reasoningaugmentedquery}.
Figure~\ref{fig:gpt4query} compares average nDCG@10 scores for four retrieval models under two query types: original queries and GPT-4 reasoning queries, each evaluated with and without \spike enhancement.
First, using \spike with original query yields performance comparable to or better than using GPT-4 reasoning queries without \spike.
Notably, this result highlights the efficiency of \spike, as it achieves similar performance to GPT-4-based query reformulation while using a much smaller 3B generator.
Second, when GPT-4 reasoning queries are used, incorporating \spike further improves retrieval performance across all models.
This consistent improvement demonstrates that \spike remains robust across various query types, including those augmented with GPT-4’s chain-of-thought reasoning. 
In other words, the scenarios generated by \spike help capture implicit relevance in reasoning-intensive tasks, regardless of the query format.

\subsection{\spike enhances retrieval experience for real-world users (RQ2)}
\label{subsec:enhancinguserex}
To examine the effectiveness of the additional information provided by \spike, we conduct a human evaluation.
Specifically, we compare different retrieved results across four criteria: \textit{Comprehensibility (Comp.)}, \textit{Specificity (Spec.)}, \textit{Usefulness (Use.)} and \textit{Overall (Over.)}.
These criteria are designed to assess different aspects of user satisfaction with the retrieved results (see Appendix~\ref{appendix:criteriahumaneval} for detailed descriptions).
Figure~\ref{fig:human_eval} presents the results of two human evaluation settings: (1) comparing standard document retrieval (Document Only) with SPIKE, and (2) comparing SPIKE with one of its variants, Document + Main Topic.
Across all evaluation criteria, \spike consistently outperforms both baselines. 
The gains are particularly notable in \textit{Specificity} and \textit{Usefulness}, where reasoning-derived information provides clearer and more practically helpful signals by explicitly expressing implicit relevance.
Furthermore, \spike offers a clear advantage in \textit{Comprehensibility}, demonstrating that users find retrieved results easier to interpret when supported by the scenario context.
Especially, the comparison between \spike and the Document + Main Topic setting (i.e., \spike w/o Explanation) highlights the critical role of the Explanation (E) component. 
While the Main Topic provides useful context, the explanation substantially enhances document comprehension and practical decision-making by explicitly expressing implicit relevance. 
\textbf{This finding clearly implies that reasoning-derived explanations significantly improve the clarity and usefulness of retrieved results for real-world users.
}

\subsection{\spike boosts RAG performance with additional context (RQ3)}
\label{subsec:effectiveRAG}
As \spike’s scenario-profiled information helps users better understand about retrieved contents, it can also serve as effective context for retrieval-augmented generation (RAG) by providing additional information alongside the retrieved content.
To verify this, we evaluate the QA performance of Claude-3.5-sonnet and Llama3.3-70B-Instruct when augmented with documents retrieved by different retrievers, comparing its performance with and without the additional context from \spike.
Specifically, we use reference answers provided in the BRIGHT benchmark and follow the evaluation process of \cite{Su2024BRIGHTAR}, where a evaluation model scores the generated answers based on their alignment with the references.
For detailed experimental settings, please refer to Appendix~\ref{appendix:RAGexperimentsetting}.
Figure~\ref{fig:rag_results} presents the average QA performance in different retrievers and generation models in the RAG setting. 
First, we observe that using documents retrieved via \spike as context consistently improves QA accuracy across all retrievers and generators.
This improvement is attributed to enhanced retrieval performance, which allows more relevant documents to be retrieved, thereby providing better context for the generator to produce accurate answers.
Moreover, further performance gains are achieved when \spike’s scenario contexts are provided alongside the retrieved documents. 
These results suggest that the scenario context not only contributes to retrieval performance but also directly enriches the retrieved content, offering stronger and more structured support for answer generation in retrieval-augmented settings.

\section{Analysis}
\label{sec:analysis}
\begin{table}[t]
    \centering
    \small
    \begin{minipage}{0.44\textwidth}
        
    
        
        \centering
        \includegraphics[width=\textwidth]{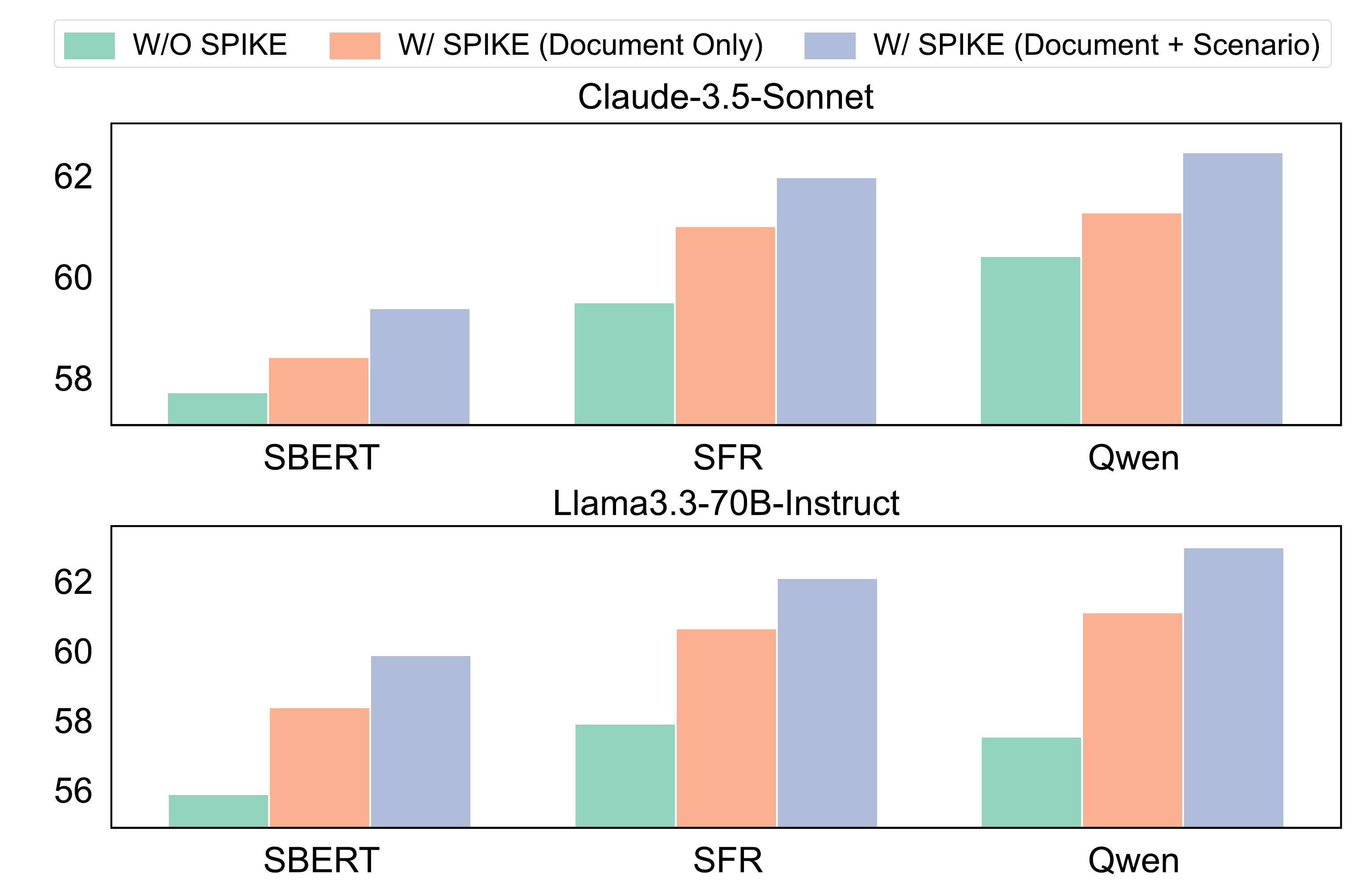}
        \captionof{figure}{Average QA performance in RAG. Document only uses retrieved documents as context; +Scenario additionally uses scenario information.}
        \label{fig:rag_results}
        
    \end{minipage}
    \hfill
    \begin{minipage}{0.54\textwidth}
        \centering
        \includegraphics[width=\textwidth]{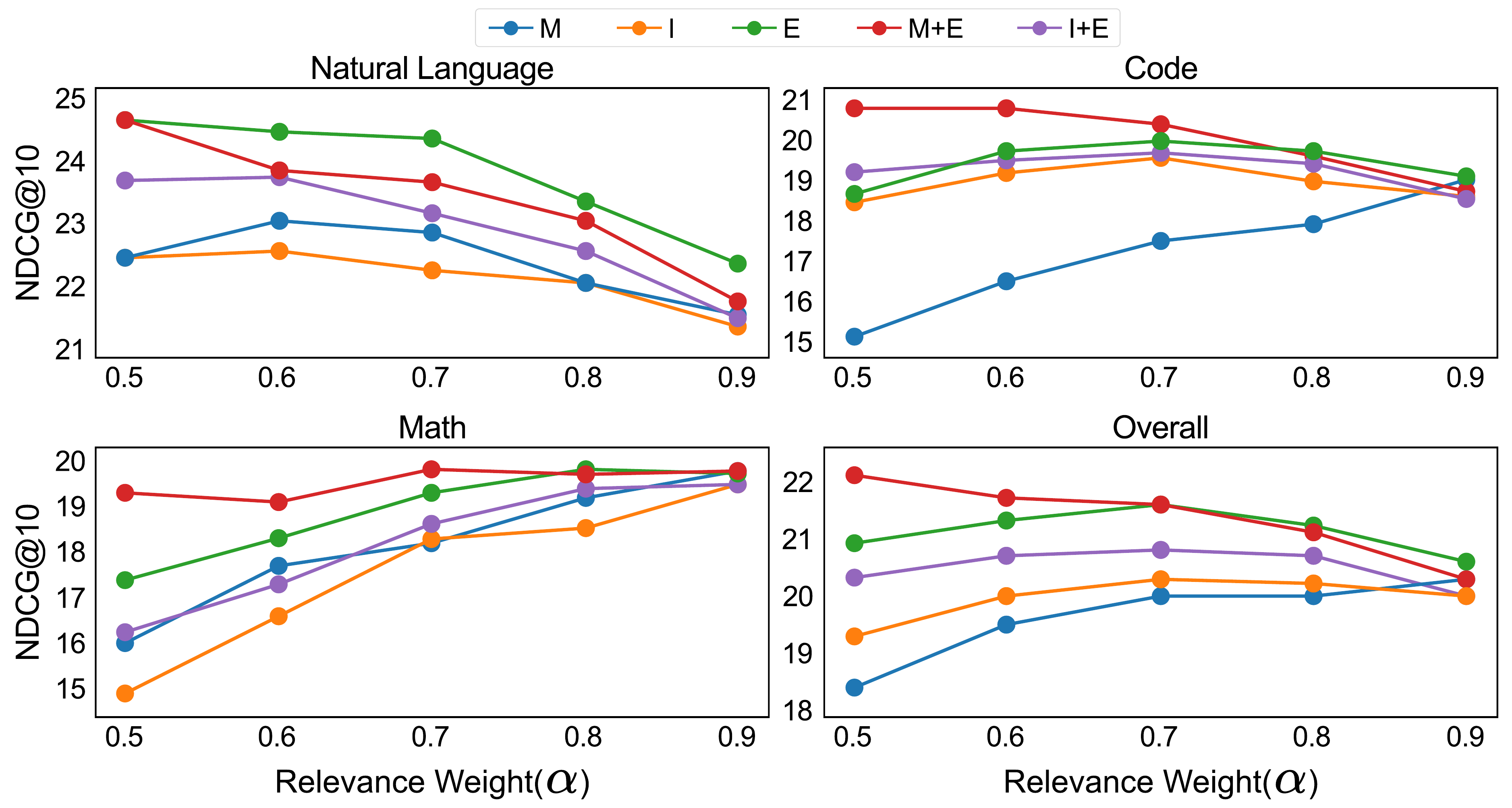}
        \captionof{figure}{Retrieval performance of \spike across different scenario components and relevance weights for each document types, using E5-Mistral-7B as the retrieval model.}
        \label{fig:component_analysis}
        
    \end{minipage}
\end{table}

\textbf{Ablation study on scenario components.}
~As shown in Figure~\ref{fig:component_analysis}, we analyze the effectiveness of different scenario-generation components within SPIKE across various document types: natural language, code, and math. 
Among the individual components (\textit{M}, \textit{I}, \textit{E}), \textit{E} achieves the highest retrieval performance, showing the importance of explicitly modeling reasoning to capture implicit relevance.
Additionally, combining \textit{E} with other components (\textit{M+E}, \textit{I+E}) led to further improvements, achieving higher retrieval performance compared to when each component was used individually. 
Notably, the combination of \textit{M+E} yielded the best overall performance, further emphasizing the benefit of combining main topic identification with reasoning-derived explanations, as discussed in Section~\ref{subsec:documentscenario}.
These results highlight that modeling implicit relevance through reasoning and indexing it for use in retrieval is more effective than relying solely on summaries or information needs.

\textbf{Ablation study on relevance weight.}
~Figure~\ref{fig:component_analysis} also illustrates the impact of the relevance weight $\alpha$ in Equation~\eqref{eq:finalrankedlist}, which balances scenario-level and document-level relevance scores. 
Across nearly all components, retrieval performance peaked at $\alpha=0.7$. 
However, different document type exhibited distinct trends. 
In natural language and code documents, M+E performance improves as $\alpha$ decreases indicating that the information provided by scenarios is more crucial than the original document content in these document types.
Conversely, in math documents, performance improves steadily as $\alpha$ increases, suggesting that document-level content (e.g., LaTeX equations) is more critical for effective retrieval. 
Given these analysis, we select $\alpha = 0.7$ as the primary configuration, as it provides the most consistent performance improvements across different types of documents.
The full results of the ablation study on scenario components and relevance weight are provided in Appendix~\ref{subsec:fullanalysis}.

\textbf{Zero-shot generalization of scenario generator.}
~To evaluate the generalizability of our scenario generator, we conduct a zero-shot experiment where the generator was trained exclusively on the BEIR corpus without exposure to the BRIGHT corpus. 
Table~\ref{tab:ood_results} presents the retrieved results under this setting. 
Notably, BEIR does not fully cover the range of domains present in BRIGHT, particularly code and math.
Despite this domain gap, the version of \spike trained only on the BEIR corpus consistently improves retrieval performance across nearly all datasets in BRIGHT. 
Surprisingly, even when trained solely on BEIR, it sometimes exhibits higher performance than when utilizing a scenario generator trained on the BRIGHT corpus (refer to Table~\ref{tab:main_results}).
Specifically, we observe substantial gains in code domain and TheoT. dataset, which are not covered by the BEIR corpus. 
These results demonstrate the strong out-of-domain generalization capability of our scenario-based retrieval framework.

\begin{table}[t!]
\setlength{\tabcolsep}{4pt}

\centering
\resizebox{\textwidth}{!}{
    \begin{tabular}{lcccccccccccccc}
    \toprule
    & \multicolumn{5}{c}{\textbf{Natural language}} 
    & \multicolumn{4}{c}{\textbf{Code}} 
    & \multicolumn{3}{c}{\textbf{Math}} 
    & \multirow{2}{*}{\centering \textbf{{Avg.}}} & \multirow{2}{*}{\centering \textbf{Improv.}} \\
    \cmidrule(r){2-6} \cmidrule(r){7-10} \cmidrule(r){11-13}
    & \textbf{Bio.} & \textbf{Earth.} & \textbf{Econ.} & \textbf{Psy.} & \textbf{Sus.} 
    & \textbf{Rob.} 
    & \textbf{Stack.} & \textbf{Leet.} & \textbf{Pony} 
    & \textbf{Aops} & \textbf{TheoQ.} & \textbf{TheoT.} \\

    \midrule
    E5-Mistral  & 18.8 & 26.0 & 15.5 & 15.8 & 18.5 & 16.4 & 9.8 & \textbf{28.7} & 4.8 & \textbf{7.1} & \textbf{26.1} & 26.8 & 17.9 & - \\
    +\spike ($C=4000$) & 25.0 & 28.5 & 17.9 & 20.0 & 20.0 & \textbf{17.7} & 14.1 & 28.6 & 9.7 & 6.1 & 23.5 & \textbf{27.9} & 19.9 & 11.5\% \\
    +\spike ($C=20000$) & \textbf{25.8} & \textbf{35.9} & \textbf{19.5} & \textbf{24.3} & \textbf{21.3} & 16.1 & \textbf{14.1} & 28.0 & \textbf{25.6} & 6.9 & 23.6 & 27.8 & \textbf{22.4} & \textbf{25.4\%} \\

    \cdashline{1-14}
    \addlinespace[2pt]
    
    SFR  & 19.5 & 26.6 & 17.8 & 19.0 & 19.8 & 16.7 & 12.7 & 27.4 & 2.0 & \textbf{7.4} & \textbf{24.3} & 26.0 & 18.3 & - \\
    +\spike ($C=4000$) & 22.0 & 29.0 & 19.0 & 25.5 & 22.0 & \textbf{17.6} & \textbf{16.2} & \textbf{27.9} & 9.6 & 6.2 & 22.7 & 26.7 & 20.4 & 11.5\% \\
    +\spike ($C=20000$) & \textbf{25.0} & \textbf{34.3} & \textbf{22.8} & \textbf{29.9} & \textbf{22.4} & 16.6 & 16.1 & 27.8 & \textbf{24.2} & 7.1 & 23.3 & \textbf{27.5} & \textbf{23.1} & \textbf{26.3\%} \\

    \cdashline{1-14}
    \addlinespace[2pt]

    GRIT &  25.0 & 32.8 & 19.0 & 19.9 & 18.0 & 17.3 & 11.6 & 29.8 & 22.0 & 8.8 & 25.1 & 21.1 & 20.9 & - \\
    +\spike ($C=4000$) & 23.0 & 27.6 & 18.4 & 20.2 & 19.5 & 18.3 & 15.8 & \textbf{31.5} & 17.5 & \textbf{8.9} & \textbf{25.6} & 25.9 & 21.0 & 0.7\% \\
    +\spike ($C=20000$) & \textbf{30.4} & \textbf{34.9} & \textbf{21.7} & \textbf{24.3} & \textbf{21.1} & \textbf{18.6} & \textbf{17.0} & 29.4 & \textbf{24.4} & 8.8 & 25.3 & \textbf{27.0} & \textbf{23.6} & \textbf{13.0\%} \\

    \cdashline{1-14}
    \addlinespace[2pt]

    Qwen & 30.9 & 36.2 & 17.7 & 24.6 & 14.9 & 13.5 & 19.9 & 25.5 & 14.4 & \textbf{27.8} & \textbf{32.9} & 32.9 & 24.3 & - \\
    +\spike ($C=4000$) & 31.3 & 40.4 & 22.8 & 24.4 & 24.5 & 16.2 & 22.8 & 24.6 & 11.1 & 13.6 & 27.8 & 33.3 & 24.4 & 0.5\% \\
    +\spike ($C=20000$) & \textbf{34.6} & \textbf{42.9} & \textbf{23.2} & \textbf{30.6} & \textbf{28.1} & \textbf{19.4} & \textbf{24.1} & \textbf{25.6} & \textbf{27.0} & 13.5 & 28.6 & \textbf{34.0} & \textbf{27.6} & \textbf{13.8\%} \\

    \bottomrule
    \end{tabular}
}
\caption{Retrieval performance when scenarios are generated by a scenario generator trained only on the BEIR corpus without the BRIGHT corpus. We report nDCG@10 for all datasets. Avg. denotes the average score across 12 datasets and Improv. denotes the improvement rate of the average score. $C$ denotes the number of documents used in the scenario-augmented training dataset. The best score on each model is shown in bold.}
\label{tab:ood_results}
\end{table}

\textbf{Ablation study on scenario-augmented training dataset.}
~To investigate the impact of scaling the scenario-augmented training dataset on retrieval performance, we conducted an ablation study on the size of the scenario-augmented training dataset used for training our scenario generator. 
Table~\ref{tab:ood_results} presents the results of this ablation study performed solely with the BEIR corpus. 
Even when utilizing only 4,000 documents, \spike demonstrates superior performance compared to the baseline without \spike.
Furthermore, scaling the scenario-augmented training dataset yields a significantly greater performance improvement. 
Notably, as previously mentioned, this performance surpasses the retrieval performance achieved when trained on the BRIGHT dataset. 
These results indicate that performance of \spike consistently improves as the scenario-augmented training dataset scales, suggesting that \spike could benefit from even larger and more diverse datasets to further enhance its capabilities to capture implicit relevance across various domains.

\textbf{Ablation study on efficient retrieval strategy.}
~To validate the effectiveness of the efficient retrieval strategy introduced in Section~\ref{subsec:efficientretrievalstrategy}, we conducted an ablation studys on the candidate set size, $k'$. 
Table~\ref{tab:efficiency} shows the retrieval performance as $k'$ is varied. 
The results indicate that performance remains nearly identical to the naive method (i.e., scoring all scenarios) as long as $k'$ is sufficiently large, while significantly reducing the computational load. 
This result demonstrates that our efficient retrieval strategy successfully provides high efficiency without compromising the retrieval effectiveness of the \spike framework.
\begin{table}[t!]
\setlength{\tabcolsep}{4pt}

\centering
\resizebox{\textwidth}{!}{
    \begin{tabular}{lccccccccccccc}
    \toprule
    & \multicolumn{5}{c}{\textbf{Natural language}} 
    & \multicolumn{4}{c}{\textbf{Code}} 
    & \multicolumn{3}{c}{\textbf{Math}} 
    & \multirow{2}{*}{\centering \textbf{{Avg.}}} \\
    \cmidrule(r){2-6} \cmidrule(r){7-10} \cmidrule(r){11-13}
    & \textbf{Bio.} & \textbf{Earth.} & \textbf{Econ.} & \textbf{Psy.} & \textbf{Sus.} 
    & \textbf{Rob.} 
    & \textbf{Stack.} & \textbf{Leet.} & \textbf{Pony} 
    & \textbf{Aops} & \textbf{TheoQ.} & \textbf{TheoT.} \\
    \midrule
    E5-Mistral  & 18.8 & 26.0 & 15.5 & 15.8 & 18.5 & 16.4 & 9.8 & 28.7 & 4.8 & 7.1 & 26.1 & 26.8 & 17.9  \\
    +\spike($k'=1000$) & 25.9 & 33.0 & 18.2 & 20.6 & 20.6 & 18.4 & 16.2 & 29.4 & 17.5 & 7.0 & 23.4 & 28.4 & 21.6 \\
    +\spike($k'=10000$) & 25.7 & 33.0 & 18.2 & 20.6 & 20.6 & 18.4 & 16.2 & 29.4 & 17.2 & 7.0 & 23.5 & 28.2 & 21.5 \\
    +\spike($k'=N$) & 25.7 & 33.0 & 18.2 & 20.6 & 20.6 & 18.4 & 16.2 & 29.4 & 17.2 & 7.0 & 23.5 & 28.2 & 21.5 \\
    
    \bottomrule
    \end{tabular}
}
\caption{Retrieval performance on BRIGHT for different values of $K'$, using E5-Mistral-7B as the retrieval model. Retrieval performance is virtually unchanged regardless of the value of $K'$, whether $K'$ is set to 1000, 10000, or $N$ (i.e., exhaustive search). }
\label{tab:efficiency}
\end{table}

\begin{figure}[!t]
    \centering
    \includegraphics[width=\columnwidth]{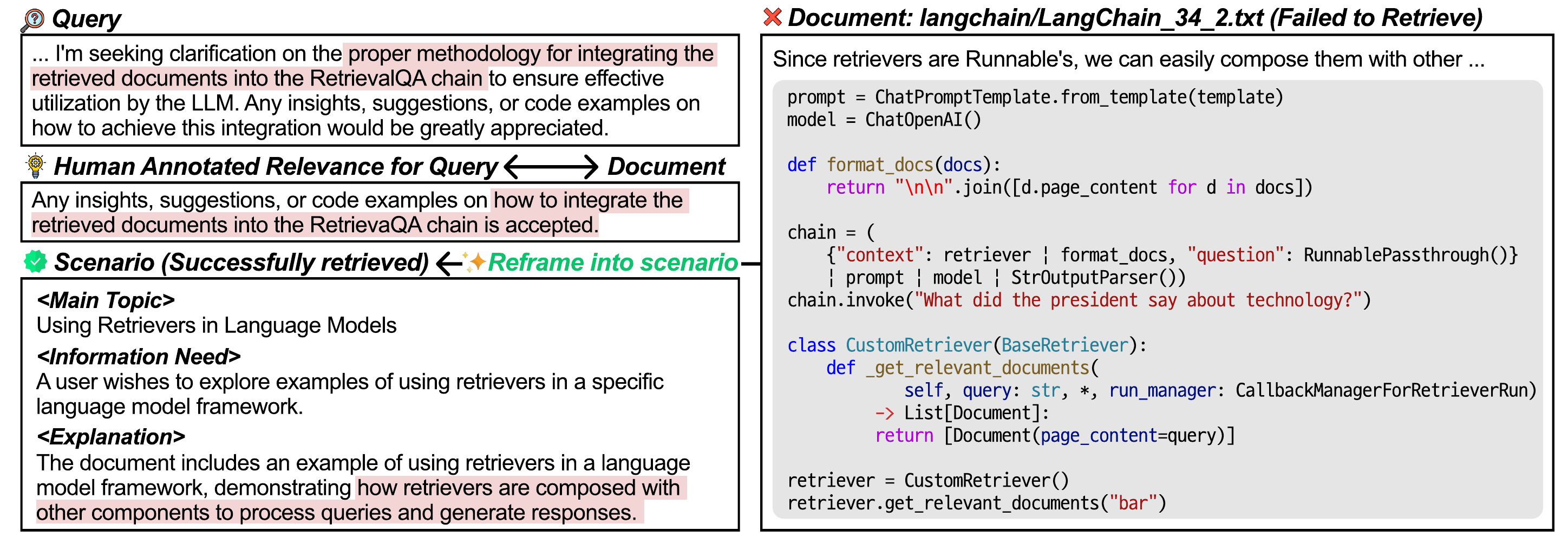}
    \caption{A StackOverflow example from BRIGHT where standard dense retrieval fails due to the absence of explicit semantic connections, while \spike retrieves the correct document by capturing implicit relevance through scenario-profiled retrieval.}
    \label{fig:casestudy}
\end{figure}

\textbf{Case study on non-natural language documents.}
~Figure~\ref{fig:casestudy} shows an example from the StackOverflow in BRIGHT. 
In this case, standard dense retrieval, which relies solely on the document, fails to retrieve the relevant result. 
In contrast, \spike successfully retrieves it.
The query seeks guidance on integrating retrieved documents into a RetrievalQA chain, but the relevant document does not explicitly contain it. 
Instead, it provides a code snippet demonstrating how retrievers interact with other components. 
Standard dense retrieval fails as it relies on surface-level similarity without reasoning, making it unable to bridge the gap between the query and the document.
\spike overcomes this limitation by leveraging scenarios that highlight latent connection between the query and the document.
This case highlights \spike’s ability to retrieve documents that require reasoning to establish implicit relevance, making it particularly effective for non-natural language content like code snippets where relevant information is not explicitly stated.

\section{Conclusion}
\label{sec:conclusion}
This paper proposes \spike, a novel dense retrieval framework designed to enhance retrieval effectiveness by explicitly modeling implicit relevance. 
By reframing documents into structured retrieval scenarios, \spike addresses the limitations of existing dense retrieval models that struggle with reasoning-intensive tasks. 
Our extensive experiments demonstrate that \spike not only enhances retrieval performance across various domains, including natural language, code, and math, but also improves usability in real-world retrieval systems and serves as an effective context for RAG-based applications. 
Furthermore, our analysis showed that \spike exhibits robust out-of-domain generalization capabilities.




\section*{Acknowledgements}
This work was supported by the IITP grants funded by the Korea government (MSIT) (No. RS-2020-II201361; RS-2024-00457882, AI Research Hub Project; IITP-2025-RS-2020-II201819).

\bibliography{reference}

\begin{thebibliography}{22}
\providecommand{\natexlab}[1]{#1}
\providecommand{\url}[1]{\texttt{#1}}
\expandafter\ifx\csname urlstyle\endcsname\relax
  \providecommand{\doi}[1]{doi: #1}\else
  \providecommand{\doi}{doi: \begingroup \urlstyle{rm}\Url}\fi

\bibitem[Boudin et~al.(2020)Boudin, Gallina, and Aizawa]{Boudin2020KeyphraseGF}
Florian Boudin, Ygor Gallina, and Akiko Aizawa.
\newblock Keyphrase generation for scientific document retrieval.
\newblock In \emph{Annual Meeting of the Association for Computational Linguistics}, 2020.
\newblock URL \url{https://api.semanticscholar.org/CorpusID:220047513}.

\bibitem[Chen et~al.(2023)Chen, Wang, Chen, Yu, Ma, Zhao, Yu, and Zhang]{Chen2023DenseXR}
Tong Chen, Hongwei Wang, Sihao Chen, Wenhao Yu, Kaixin Ma, Xinran Zhao, Dong Yu, and Hongming Zhang.
\newblock Dense x retrieval: What retrieval granularity should we use?
\newblock In \emph{Conference on Empirical Methods in Natural Language Processing}, 2023.
\newblock URL \url{https://api.semanticscholar.org/CorpusID:266163052}.

\bibitem[Chen et~al.(2024)Chen, Yoon, Sachan, Wang, Cohen-Addad, Bateni, Lee, and Pfister]{Chen2024ReInvokeTI}
Yanfei Chen, Jinsung Yoon, Devendra~Singh Sachan, Qingze Wang, Vincent Cohen-Addad, MohammadHossein Bateni, Chen-Yu Lee, and Tomas Pfister.
\newblock Re-invoke: Tool invocation rewriting for zero-shot tool retrieval.
\newblock In \emph{Conference on Empirical Methods in Natural Language Processing}, 2024.
\newblock URL \url{https://api.semanticscholar.org/CorpusID:271709437}.

\bibitem[Cheriton(2019)]{Cheriton2019FromDT}
David~R. Cheriton.
\newblock From doc2query to doctttttquery.
\newblock 2019.
\newblock URL \url{https://api.semanticscholar.org/CorpusID:208612557}.

\bibitem[Dubey et~al.(2024)Dubey, Jauhri, Pandey, Kadian, Al-Dahle, Letman, Mathur, Schelten, Yang, Fan, Goyal, Hartshorn, Yang, Mitra, Sravankumar, Korenev, Hinsvark, Rao, Zhang, Rodriguez, Gregerson, Spataru, tiste Roziere, Biron, Tang, Chern, Caucheteux, Nayak, Bi, Marra, McConnell, Keller, Touret, Wu, Wong, Ferrer, Nikolaidis, Allonsius, Song, Pintz, Livshits, Esiobu, Choudhary, Mahajan, Garcia-Olano, Perino, Hupkes, Lakomkin, AlBadawy, Lobanova, Dinan, Smith, Radenovic, Zhang, Synnaeve, Lee, Anderson, Nail, Mialon, Pang, Cucurell, Nguyen, Korevaar, Xu, Touvron, Zarov, Ibarra, Kloumann, Misra, Evtimov, Copet, Lee, Geffert, Vranes, Park, Mahadeokar, Shah, van~der Linde, Billock, Hong, Lee, Fu, Chi, Huang, Liu, Wang, Yu, Bitton, Spisak, Park, Rocca, Johnstun, Saxe, Jia, Alwala, Upasani, Plawiak, Li, neth Heafield, Stone, El-Arini, Iyer, Malik, ley Chiu, Bhalla, Rantala-Yeary, van~der Maaten, Chen, Tan, Jenkins, Martin, Madaan, Malo, Blecher, Landzaat, de~Oliveira, Muzzi, Pasupuleti, Singh, Paluri, Kardas,
  Oldham, Rita, Pavlova, Kambadur, Lewis, Si, Singh, Hassan, Goyal, Torabi, Bashlykov, Bogoychev, Chatterji, Duchenne, cCelebi, Alrassy, Zhang, Li, Vasi{\'c}, Weng, Bhargava, Dubal, Krishnan, Koura, Xu, He, Dong, Srinivasan, Ganapathy, Calderer, Cabral, Stojnic, Raileanu, Girdhar, Patel, main Sauvestre, Polidoro, Sumbaly, Taylor, Silva, Hou, Wang, Hosseini, Chennabasappa, Singh, Bell, Kim, Edunov, Nie, Narang, Raparthy, Shen, Wan, Bhosale, Zhang, Vandenhende, Batra, Whitman, Sootla, Collot, Gururangan, Borodinsky, Herman, Fowler, Sheasha, Georgiou, Scialom, Speckbacher, Mihaylov, Xiao, Karn, Goswami, Gupta, Ramanathan, Kerkez, Gonguet, Do, Vogeti, Petrovic, Chu, Xiong, Fu, ney Meers, Martinet, Wang, Tan, Xie, Jia, Wang, Goldschlag, Gaur, Babaei, Wen, Song, Zhang, Li, Mao, Coudert, Yan, Chen, Papakipos, Singh, Grattafiori, Jain, Kelsey, Shajnfeld, Gangidi, Victoria, Goldstand, Menon, Sharma, Boesenberg, Vaughan, Baevski, Feinstein, Kallet, Sangani, Yunus, Lupu, Alvarado, Caples, Gu, Ho, Poulton, Ryan,
  Ramchandani, Franco, Saraf, Chowdhury, Gabriel, Bharambe, Eisenman, Yazdan, James, Maurer, Leonhardi, Huang, Loyd, Paola, Paranjape, Liu, Wu, Ni, Hancock, Wasti, Spence, Stojkovic, Gamido, Montalvo, Parker, Burton, Mejia, Wang, Kim, Zhou, Hu, Chu, Cai, Tindal, Feichtenhofer, Civin, Beaty, Kreymer, Li, Wyatt, Adkins, Xu, Testuggine, David, Parikh, Liskovich, Foss, Wang, Le, Holland, Dowling, Jamil, Montgomery, Presani, Hahn, Wood, Brinkman, Arcaute, Dunbar, Smothers, Sun, Kreuk, Tian, Ozgenel, Caggioni, Guzm'an, Kanayet, Seide, Florez, Schwarz, Badeer, Swee, Halpern, Thattai, Herman, Sizov, Zhang, Lakshminarayanan, Shojanazeri, Zou, Wang, Zha, Habeeb, Rudolph, Suk, Aspegren, Goldman, Molybog, Tufanov, Veliche, Gat, Weissman, Geboski, Kohli, Asher, Gaya, Marcus, Tang, Chan, Zhen, Reizenstein, Teboul, Zhong, Jin, Yang, Cummings, Carvill, Shepard, McPhie, Torres, Ginsburg, Wang, Wu, KamHou, Saxena, Prasad, Khandelwal, Zand, Matosich, Veeraraghavan, Michelena, Li, Huang, Chawla, Lakhotia, Huang, Chen, Garg,
  Lavender, Silva, Bell, Zhang, Guo, Yu, Moshkovich, Wehrstedt, Khabsa, Avalani, Bhatt, Tsimpoukelli, Mankus, Hasson, Lennie, Reso, Groshev, Naumov, Lathi, Keneally, Seltzer, Valko, Restrepo, Patel, Vyatskov, Samvelyan, Clark, Macey, Wang, Hermoso, Metanat, Rastegari, Bansal, Santhanam, Parks, White, Bawa, Singhal, Egebo, Usunier, Laptev, Dong, Zhang, Cheng, Chernoguz, Hart, Salpekar, Kalinli, Kent, Parekh, Saab, Balaji, Rittner, Bontrager, Roux, Doll{\'a}r, Zvyagina, Ratanchandani, Yuvraj, Liang, Alao, Rodriguez, Ayub, Murthy, Nayani, Mitra, Li, Hogan, Battey, Wang, Maheswari, Howes, Rinott, Bondu, Datta, Chugh, Hunt, Dhillon, Sidorov, Pan, Verma, Yamamoto, Ramaswamy, Lindsay, Feng, Lin, Zha, Shankar, Zhang, Wang, Agarwal, Sajuyigbe, Chintala, Max, Chen, Kehoe, Satterfield, Govindaprasad, Gupta, Cho, Virk, Subramanian, Choudhury, Goldman, Remez, Glaser, Best, Kohler, Robinson, Li, Zhang, Matthews, Chou, Shaked, Vontimitta, Ajayi, Montanez, Mohan, Kumar, Mangla, Ionescu, Poenaru, Mihailescu, Ivanov, Li, Wang,
  Jiang, Bouaziz, Constable, Tang, Wang, Wu, Wang, Xia, Wu, Gao, Chen, Hu, Jia, Qi, Li, Zhang, Zhang, Adi, Nam, Wang, Hao, Qian, He, Rait, DeVito, Rosnbrick, Wen, Yang, and Zhao]{Dubey2024TheL3}
Abhimanyu Dubey, Abhinav Jauhri, Abhinav Pandey, Abhishek Kadian, Ahmad Al-Dahle, Aiesha Letman, Akhil Mathur, Alan Schelten, Amy Yang, Angela Fan, Anirudh Goyal, Anthony~S. Hartshorn, Aobo Yang, Archi Mitra, Archie Sravankumar, Artem Korenev, Arthur Hinsvark, Arun Rao, Aston Zhang, Aur{\'e}lien Rodriguez, Austen Gregerson, Ava Spataru, Bap tiste Roziere, Bethany Biron, Binh Tang, Bobbie Chern, Charlotte Caucheteux, Chaya Nayak, Chloe Bi, Chris Marra, Chris McConnell, Christian Keller, Christophe Touret, Chunyang Wu, Corinne Wong, Cristian~Cant{\'o}n Ferrer, Cyrus Nikolaidis, Damien Allonsius, Daniel Song, Danielle Pintz, Danny Livshits, David Esiobu, Dhruv Choudhary, Dhruv Mahajan, Diego Garcia-Olano, Diego Perino, Dieuwke Hupkes, Egor Lakomkin, Ehab~A. AlBadawy, Elina Lobanova, Emily Dinan, Eric~Michael Smith, Filip Radenovic, Frank Zhang, Gabriele Synnaeve, Gabrielle Lee, Georgia~Lewis Anderson, Graeme Nail, Gr{\'e}goire Mialon, Guanglong Pang, Guillem Cucurell, Hailey Nguyen, Hannah Korevaar, Hu~Xu, Hugo
  Touvron, Iliyan Zarov, Imanol~Arrieta Ibarra, Isabel~M. Kloumann, Ishan Misra, Ivan Evtimov, Jade Copet, Jaewon Lee, Jan~Laurens Geffert, Jana Vranes, Jason Park, Jay Mahadeokar, Jeet Shah, Jelmer van~der Linde, Jennifer Billock, Jenny Hong, Jenya Lee, Jeremy Fu, Jianfeng Chi, Jianyu Huang, Jiawen Liu, Jie Wang, Jiecao Yu, Joanna Bitton, Joe Spisak, Jongsoo Park, Joseph Rocca, Joshua Johnstun, Joshua Saxe, Ju-Qing Jia, Kalyan~Vasuden Alwala, K.~Upasani, Kate Plawiak, Keqian Li, Ken-591 neth Heafield, Kevin Stone, Khalid El-Arini, Krithika Iyer, Kshitiz Malik, Kuen ley Chiu, Kunal Bhalla, Lauren Rantala-Yeary, Laurens van~der Maaten, Lawrence Chen, Liang Tan, Liz Jenkins, Louis Martin, Lovish Madaan, Lubo Malo, Lukas Blecher, Lukas Landzaat, Luke de~Oliveira, Madeline Muzzi, Mahesh~Babu Pasupuleti, Mannat Singh, Manohar Paluri, Marcin Kardas, Mathew Oldham, Mathieu Rita, Maya Pavlova, Melissa Hall~Melanie Kambadur, Mike Lewis, Min Si, Mitesh~Kumar Singh, Mona Hassan, Naman Goyal, Narjes Torabi, Nikolay
  Bashlykov, Nikolay Bogoychev, Niladri~S. Chatterji, Olivier Duchenne, Onur cCelebi, Patrick Alrassy, Pengchuan Zhang, Pengwei Li, Petar Vasi{\'c}, Peter Weng, Prajjwal Bhargava, Pratik Dubal, Praveen Krishnan, Punit~Singh Koura, Puxin Xu, Qing He, Qingxiao Dong, Ragavan Srinivasan, Raj Ganapathy, Ramon Calderer, Ricardo~Silveira Cabral, Robert Stojnic, Roberta Raileanu, Rohit Girdhar, Rohit Patel, Ro~main Sauvestre, Ronnie Polidoro, Roshan Sumbaly, Ross Taylor, Ruan Silva, Rui Hou, Rui Wang, Saghar Hosseini, Sahana Chennabasappa, Sanjay Singh, Sean Bell, Seohyun~Sonia Kim, Sergey Edunov, Shaoliang Nie, Sharan Narang, Sharath~Chandra Raparthy, Sheng Shen, Shengye Wan, Shruti Bhosale, Shun Zhang, Simon Vandenhende, Soumya Batra, Spencer Whitman, Sten Sootla, Stephane Collot, Suchin Gururangan, Sydney Borodinsky, Tamar Herman, Tara Fowler, Tarek Sheasha, Thomas Georgiou, Thomas Scialom, Tobias Speckbacher, Todor Mihaylov, Tong Xiao, Ujjwal Karn, Vedanuj Goswami, Vibhor Gupta, Vignesh Ramanathan, Viktor Kerkez,
  Vincent Gonguet, Virginie Do, Vish Vogeti, Vladan Petrovic, Weiwei Chu, Wenhan Xiong, Wenyin Fu, Whit ney Meers, Xavier Martinet, Xiaodong Wang, Xiaoqing~Ellen Tan, Xinfeng Xie, Xuchao Jia, Xuewei Wang, Yaelle Goldschlag, Yashesh Gaur, Yasmine Babaei, Yiqian Wen, Yiwen Song, Yuchen Zhang, Yue Li, Yuning Mao, Zacharie~Delpierre Coudert, Zhengxu Yan, Zhengxing Chen, Zoe Papakipos, Aaditya~K. Singh, Aaron Grattafiori, Abha Jain, Adam Kelsey, Adam Shajnfeld, Adi Gangidi, Adolfo Victoria, Ahuva Goldstand, Ajay Menon, Ajay Sharma, Alex Boesenberg, Alex Vaughan, Alexei Baevski, Allie Feinstein, Amanda Kallet, Amit Sangani, Anam Yunus, Andrei Lupu, Andres Alvarado, Andrew Caples, Andrew Gu, Andrew Ho, Andrew Poulton, Andrew Ryan, Ankit Ramchandani, Annie Franco, Aparajita Saraf, Arkabandhu Chowdhury, Ashley Gabriel, Ashwin Bharambe, Assaf Eisenman, Azadeh Yazdan, Beau James, Ben Maurer, Ben Leonhardi, Po-Yao~(Bernie) Huang, Beth Loyd, Beto~De Paola, Bhargavi Paranjape, Bing Liu, Bo~Wu, Boyu Ni, Braden Hancock, Bram
  Wasti, Brandon Spence, Brani Stojkovic, Brian Gamido, Britt Montalvo, Carl Parker, Carly Burton, Catalina Mejia, Changhan Wang, Changkyu Kim, Chao Zhou, Chester Hu, Ching-Hsiang Chu, Chris Cai, Chris Tindal, Christoph Feichtenhofer, Damon Civin, Dana Beaty, Daniel Kreymer, Shang-Wen Li, Danny Wyatt, David Adkins, David Xu, Davide Testuggine, Delia David, Devi Parikh, Diana Liskovich, Didem Foss, Dingkang Wang, Duc Le, Dustin Holland, Edward Dowling, Eissa Jamil, Elaine Montgomery, Eleonora Presani, Emily Hahn, Emily Wood, Erik Brinkman, Esteban Arcaute, Evan Dunbar, Evan Smothers, Fei Sun, Felix Kreuk, Feng Tian, Firat Ozgenel, Francesco Caggioni, Francisco Guzm'an, Frank~J. Kanayet, Frank Seide, Gabriela~Medina Florez, Gabriella Schwarz, Gada Badeer, Georgia Swee, Gil Halpern, Govind Thattai, Grant Herman, Grigory~G. Sizov, Guangyi Zhang, Guna Lakshminarayanan, Hamid Shojanazeri, Han Zou, Hannah Wang, Han Zha, Haroun Habeeb, Harrison Rudolph, Helen Suk, Henry Aspegren, Hunter Goldman, Igor Molybog, Igor
  Tufanov, Irina-Elena Veliche, Itai Gat, Jake Weissman, James Geboski, James Kohli, Japhet Asher, Jean-Baptiste Gaya, Jeff Marcus, Jeff Tang, Jennifer Chan, Jenny Zhen, Jeremy Reizenstein, Jeremy Teboul, Jessica Zhong, Jian Jin, Jingyi Yang, Joe Cummings, Jon Carvill, Jon Shepard, Jonathan McPhie, Jonathan Torres, Josh Ginsburg, Junjie Wang, Kaixing(Kai) Wu, U~KamHou, Karan Saxena, Karthik Prasad, Kartikay Khandelwal, Katayoun Zand, Kathy Matosich, Kaushik Veeraraghavan, Kelly Michelena, Keqian Li, Kun Huang, Kunal Chawla, Kushal Lakhotia, Kyle Huang, Lailin Chen, Lakshya Garg, A~Lavender, Leandro Silva, Lee Bell, Lei Zhang, Liangpeng Guo, Licheng Yu, Liron Moshkovich, Luca Wehrstedt, Madian Khabsa, Manav Avalani, Manish Bhatt, Maria Tsimpoukelli, Martynas Mankus, Matan Hasson, Matthew Lennie, Matthias Reso, Maxim Groshev, Maxim Naumov, Maya Lathi, Meghan Keneally, Michael~L. Seltzer, Michal Valko, Michelle Restrepo, Mihir Patel, Mik Vyatskov, Mikayel Samvelyan, Mike Clark, Mike Macey, Mike Wang,
  Miquel~Jubert Hermoso, Mo~Metanat, Mohammad Rastegari, Munish Bansal, Nandhini Santhanam, Natascha Parks, Natasha White, Navyata Bawa, Nayan Singhal, Nick Egebo, Nicolas Usunier, Nikolay~Pavlovich Laptev, Ning Dong, Ning Zhang, Norman Cheng, Oleg Chernoguz, Olivia Hart, Omkar Salpekar, Ozlem Kalinli, Parkin Kent, Parth Parekh, Paul Saab, Pavan Balaji, Pedro Rittner, Philip Bontrager, Pierre Roux, Piotr Doll{\'a}r, Polina Zvyagina, Prashant Ratanchandani, Pritish Yuvraj, Qian Liang, Rachad Alao, Rachel Rodriguez, Rafi Ayub, Raghotham Murthy, Raghu Nayani, Rahul Mitra, Raymond Li, Rebekkah Hogan, Robin Battey, Rocky Wang, Rohan Maheswari, Russ Howes, Ruty Rinott, Sai~Jayesh Bondu, Samyak Datta, Sara Chugh, Sara Hunt, Sargun Dhillon, Sasha Sidorov, Satadru Pan, Saurabh Verma, Seiji Yamamoto, Sharadh Ramaswamy, Shaun Lindsay, Sheng Feng, Shenghao Lin, Shengxin~Cindy Zha, Shiva Shankar, Shuqiang Zhang, Sinong Wang, Sneha Agarwal, Soji Sajuyigbe, Soumith Chintala, Stephanie Max, Stephen Chen, Steve Kehoe, Steve
  Satterfield, Sudarshan Govindaprasad, Sumit Gupta, Sung-Bae Cho, Sunny Virk, Suraj Subramanian, Sy~Choudhury, Sydney Goldman, Tal Remez, Tamar Glaser, Tamara Best, Thilo Kohler, Thomas Robinson, Tianhe Li, Tianjun Zhang, Tim Matthews, Timothy Chou, Tzook Shaked, Varun Vontimitta, Victoria Ajayi, Victoria Montanez, Vijai Mohan, Vinay~Satish Kumar, Vishal Mangla, Vlad Ionescu, Vlad~Andrei Poenaru, Vlad~T. Mihailescu, Vladimir Ivanov, Wei Li, Wenchen Wang, Wenwen Jiang, Wes Bouaziz, Will Constable, Xia Tang, Xiaofang Wang, Xiaojian Wu, Xiaolan Wang, Xide Xia, Xilun Wu, Xinbo Gao, Yanjun Chen, Ye~Hu, Ye~Jia, Ye~Qi, Yenda Li, Yilin Zhang, Ying Zhang, Yossi Adi, Youngjin Nam, Yu~Wang, Yuchen Hao, Yundi Qian, Yuzi He, Zach Rait, Zachary DeVito, Zef Rosnbrick, Zhaoduo Wen, Zhenyu Yang, and Zhiwei Zhao.
\newblock The llama 3 herd of models.
\newblock \emph{ArXiv}, abs/2407.21783, 2024.
\newblock URL \url{https://api.semanticscholar.org/CorpusID:271571434}.

\bibitem[Hu et~al.(2021)Hu, Shen, Wallis, Allen-Zhu, Li, Wang, and Chen]{Hu2021LoRALA}
J.~Edward Hu, Yelong Shen, Phillip Wallis, Zeyuan Allen-Zhu, Yuanzhi Li, Shean Wang, and Weizhu Chen.
\newblock Lora: Low-rank adaptation of large language models.
\newblock \emph{ArXiv}, abs/2106.09685, 2021.
\newblock URL \url{https://api.semanticscholar.org/CorpusID:235458009}.

\bibitem[Jeong et~al.(2021)Jeong, Baek, chaeHun Park, and Park]{Jeong2021UnsupervisedDE}
Soyeong Jeong, Jinheon Baek, chaeHun Park, and Jong~C. Park.
\newblock Unsupervised document expansion for information retrieval with stochastic text generation.
\newblock \emph{ArXiv}, abs/2105.00666, 2021.
\newblock URL \url{https://api.semanticscholar.org/CorpusID:233481135}.

\bibitem[Karpukhin et~al.(2020)Karpukhin, Oguz, Min, Lewis, Wu, Edunov, Chen, and Yih]{EMNLP/KarpukhinOMLWEC20/DPR}
Vladimir Karpukhin, Barlas Oguz, Sewon Min, Patrick S.~H. Lewis, Ledell Wu, Sergey Edunov, Danqi Chen, and Wen{-}tau Yih.
\newblock Dense passage retrieval for open-domain question answering.
\newblock In \emph{EMNLP}, pp.\  6769--6781, 2020.
\newblock URL \url{https://doi.org/10.18653/v1/2020.emnlp-main.550}.

\bibitem[Khattab \& Zaharia(2020)Khattab and Zaharia]{sigir/KhattabZ20/ColBERT}
Omar Khattab and Matei Zaharia.
\newblock Colbert: Efficient and effective passage search via contextualized late interaction over {BERT}.
\newblock In \emph{SIGIR}, pp.\  39--48, 2020.
\newblock URL \url{https://doi.org/10.1145/3397271.3401075}.

\bibitem[Li et~al.(2023)Li, Zhang, Zhang, Long, Xie, and Zhang]{li2023towards}
Zehan Li, Xin Zhang, Yanzhao Zhang, Dingkun Long, Pengjun Xie, and Meishan Zhang.
\newblock Towards general text embeddings with multi-stage contrastive learning.
\newblock \emph{arXiv preprint arXiv:2308.03281}, 2023.

\bibitem[Meng et~al.(2024)Meng, Liu, Joty, Xiong, Zhou, and Yavuz]{meng2024sfrembedding}
Rui Meng, Ye~Liu, Shafiq~Rayhan Joty, Caiming Xiong, Yingbo Zhou, and Semih Yavuz.
\newblock Sfrembedding-mistral: enhance text retrieval with transfer learning.
\newblock \emph{Salesforce AI Research Blog}, 3, 2024.

\bibitem[Muennighoff et~al.(2022)Muennighoff, Tazi, Magne, and Reimers]{Muennighoff2022MTEBMT}
Niklas Muennighoff, Nouamane Tazi, Loic Magne, and Nils Reimers.
\newblock Mteb: Massive text embedding benchmark.
\newblock In \emph{Conference of the European Chapter of the Association for Computational Linguistics}, 2022.
\newblock URL \url{https://api.semanticscholar.org/CorpusID:252907685}.

\bibitem[Muennighoff et~al.(2024)Muennighoff, Su, Wang, Yang, Wei, Yu, Singh, and Kiela]{muennighoff2024GRIT}
Niklas Muennighoff, Hongjin Su, Liang Wang, Nan Yang, Furu Wei, Tao Yu, Amanpreet Singh, and Douwe Kiela.
\newblock Generative representational instruction tuning.
\newblock \emph{arXiv preprint arXiv:2402.09906}, 2024.

\bibitem[Niu et~al.(2024)Niu, Joty, Liu, Xiong, Zhou, and Yavuz]{Niu2024JudgeRankLL}
Tong Niu, Shafiq Joty, Ye~Liu, Caiming Xiong, Yingbo Zhou, and Semih Yavuz.
\newblock Judgerank: Leveraging large language models for reasoning-intensive reranking.
\newblock \emph{ArXiv}, abs/2411.00142, 2024.
\newblock URL \url{https://api.semanticscholar.org/CorpusID:273798418}.

\bibitem[Nogueira et~al.(2019)Nogueira, Yang, Lin, and Cho]{Nogueira2019DocumentEB}
Rodrigo Nogueira, Wei Yang, Jimmy~J. Lin, and Kyunghyun Cho.
\newblock Document expansion by query prediction.
\newblock \emph{ArXiv}, abs/1904.08375, 2019.
\newblock URL \url{https://api.semanticscholar.org/CorpusID:119314259}.

\bibitem[Reimers \& Gurevych(2019)Reimers and Gurevych]{reimers2019sentence}
Nils Reimers and Iryna Gurevych.
\newblock Sentence-bert: Sentence embeddings using siamese bert-networks.
\newblock \emph{arXiv preprint arXiv:1908.10084}, 2019.

\bibitem[Sarthi et~al.(2024)Sarthi, Abdullah, Tuli, Khanna, Goldie, and Manning]{Sarthi2024RAPTORRA}
Parth Sarthi, Salman Abdullah, Aditi Tuli, Shubh Khanna, Anna Goldie, and Christopher~D. Manning.
\newblock Raptor: Recursive abstractive processing for tree-organized retrieval.
\newblock \emph{ArXiv}, abs/2401.18059, 2024.
\newblock URL \url{https://api.semanticscholar.org/CorpusID:267334785}.

\bibitem[Su et~al.(2024)Su, Yen, Xia, Shi, Muennighoff, yu~Wang, Liu, Shi, Siegel, Tang, Sun, Yoon, Arik, Chen, and Yu]{Su2024BRIGHTAR}
Hongjin Su, Howard Yen, Mengzhou Xia, Weijia Shi, Niklas Muennighoff, Han yu~Wang, Haisu Liu, Quan Shi, Zachary~S. Siegel, Michael Tang, Ruoxi Sun, Jinsung Yoon, Sercan~{\"O}. Arik, Danqi Chen, and Tao Yu.
\newblock Bright: A realistic and challenging benchmark for reasoning-intensive retrieval.
\newblock \emph{ArXiv}, abs/2407.12883, 2024.
\newblock URL \url{https://api.semanticscholar.org/CorpusID:271270735}.

\bibitem[Thakur et~al.(2021)Thakur, Reimers, Ruckl'e, Srivastava, and Gurevych]{Thakur2021BEIRAH}
Nandan Thakur, Nils Reimers, Andreas Ruckl'e, Abhishek Srivastava, and Iryna Gurevych.
\newblock Beir: A heterogenous benchmark for zero-shot evaluation of information retrieval models.
\newblock \emph{ArXiv}, abs/2104.08663, 2021.
\newblock URL \url{https://api.semanticscholar.org/CorpusID:233296016}.

\bibitem[Wang et~al.(2023)Wang, Yang, Huang, Yang, Majumder, and Wei]{Wang2023ImprovingTE}
Liang Wang, Nan Yang, Xiaolong Huang, Linjun Yang, Rangan Majumder, and Furu Wei.
\newblock Improving text embeddings with large language models.
\newblock \emph{ArXiv}, abs/2401.00368, 2023.
\newblock URL \url{https://api.semanticscholar.org/CorpusID:266693831}.

\bibitem[Xiao et~al.(2023)Xiao, Liu, Zhang, and Muennighoff]{bge_embedding}
Shitao Xiao, Zheng Liu, Peitian Zhang, and Niklas Muennighoff.
\newblock C-pack: Packaged resources to advance general chinese embedding, 2023.

\bibitem[Zhang et~al.(2019)Zhang, Zhao, Saleh, and Liu]{zhang2019pegasus}
Jingqing Zhang, Yao Zhao, Mohammad Saleh, and Peter~J. Liu.
\newblock Pegasus: Pre-training with extracted gap-sentences for abstractive summarization, 2019.

\end{thebibliography}
\bibliographystyle{colm2025_conference}

\newpage
\DoToC
\newpage

\appendix
\section{Comparison with existing document expansion works}
\label{sec:comparisonwithexisting}
\subsection{Experimental settings}
As discussed in Section~\ref{sec:relatedworks}, prior works have improved document representations by expanding them with pseudo queries~\citep{Nogueira2019DocumentEB, Chen2024ReInvokeTI} or summaries~\citep{Jeong2021UnsupervisedDE}.
To evaluate whether \spike offers a more effective approach in reasoning-intensive retrieval tasks, we conduct additional comparative experiments against these traditional document expansion methods.
However, directly comparing \spike to existing methods is not entirely fair, as prior approaches typically rely on models such as DocT5Query~\citep{Cheriton2019FromDT} or PEGASUS-Large~\citep{zhang2019pegasus}, which not only lack the ability to handle non-natural language documents like code but also fall significantly short in overall language understanding capabilities compared to the LLM used in \spike.
To ensure a fair comparison, we use the same LLM backbone (Llama3.2-3B-Instruct) that is used for \spike’s scenario generator, and apply it to generate pseudo queries and summary for each document.
Specifically, we generate three pseudo queries for the pseudo query document expansion and one summary for the summary document expansion per document, and concatenate them with the document representation.

\subsection{Results}
Table~\ref{tab:compareexisting} presents the performance comparison between \spike, pseudo query-based document expansion, and summary-based document expansion.
First, \spike consistently improves average performance across all dense retrievers, whereas traditional document expansion methods leveraging pseudo queries or summary yield smaller gains compared to \spike, and even lead to performance degradation.
In particular, document expansion approach that leveraging pseudo queries fails to provide any improvements and even degrades performance across all dense retrievers.
These results suggest that the additional context used in prior document expansion methods may introduce noise, especially in reasoning-intensive retrieval tasks where surface-level semantic information is insufficient. 

\begin{table}[h!]
\setlength{\tabcolsep}{4pt}
\centering

\resizebox{\textwidth}{!}{
    \begin{tabular}{lcccccccccccccc}
    \toprule
    & \multicolumn{5}{c}{\textbf{Natural language}} 
    & \multicolumn{4}{c}{\textbf{Code}} 
    & \multicolumn{3}{c}{\textbf{Math}} 
    & \multirow{2}{*}{\centering \textbf{{Avg.}}} & \multirow{2}{*}{\centering \textbf{{Improv.}}} \\
    \cmidrule(r){2-6} \cmidrule(r){7-10} \cmidrule(r){11-13}
    & \textbf{Bio.} & \textbf{Earth.} & \textbf{Econ.} & \textbf{Psy.} & \textbf{Sus.} 
    & \textbf{Rob.} 
    & \textbf{Stack.} & \textbf{Leet.} & \textbf{Pony} 
    & \textbf{Aops} & \textbf{TheoQ.} & \textbf{TheoT.} \\
    \midrule
    E5-Mistral 
    & 18.8 & 26.0 & 15.5 & 15.8 & 18.5
    & 16.4 & 9.8 & 28.7 & 4.8
    & \textbf{7.1} & \textbf{26.1} & 26.8
    & 17.9 & - \\

    +PQ.
    & 18.6 &28.3 &13.2 &12.7 & 15.4
    & 14.6 & 14.3 & 29.9 & 3.3
    & 4.8 & 19.7 & 16.5
    & 15.9 & -11.2\% \\
    
    +Sum.
    & 17.8 & 24.9 & 16.5 & 18.5 & \textbf{22.6}
    & 16.3 & 15.8 & \textbf{31.3} & 0.6 
    & 5.3 & 22.5 & 19.8
    & 17.7 & -1.1\% \\

    +\spike
    & \textbf{25.9} & \textbf{33.0} & \textbf{18.2} & \textbf{20.6} & {20.6}
    & \textbf{18.4} & \textbf{16.2} & {29.4} & \textbf{17.5}
    & 7.0 & 23.4 & \textbf{28.4}
    & \textbf{21.6} & \textbf{+20.7\%} \\

    \midrule
    SFR
    & 19.5 & 26.6 & 17.8 & 19.0 & 19.8
    & 16.7
    & 12.7 & 27.4 & 2.0
    & \textbf{7.4} & \textbf{24.3} & 26.0
    & 18.3 & - \\

    +PQ.
    & 17.9 & 28.5 & 16.2 & 17.7 & 17.6
    & 15.1 & 15.8 & 29.4 & 3.7
    & 5.8 & 19.6 & 19.2
    & 17.2 & -6.0\% \\
    
    +Sum.
    & 20.6 & 27.6 & 18.5 & 23.3 & \textbf{24.6}
    & 16.7 & 17.4 & \textbf{31.6} & 1.1
    & 5.6 & 22.7 & 22.3
    & 19.3 & 5.5\% \\

    +\spike 
    & \textbf{23.6} & \textbf{31.7} & \textbf{19.9} & \textbf{26.0} & {21.2}
    & \textbf{17.8} & \textbf{17.6} & 28.6 & \textbf{17.3}
    & 6.5 & 22.8 & \textbf{27.5}
    & \textbf{21.7} & \textbf{+18.6\%} \\

    \midrule
    GRIT
    & 25.0 & \textbf{32.8} & 19.0 & 19.9 & 18.0
    & 17.3 & 11.6 & 29.8 & \textbf{22.0}
    & 8.8 & 25.1 & 21.1
    & 20.9 & - \\

    +PQ.
    & 20.3 & 24.6 & 16.0 & 18.9 & 18.5
    & 18.8 & 13.3 & \textbf{33.4} & 7.6
    & 7.4 & 23.4 & 19.2
    & 18.4 & -12.0\% \\
    
    +Sum.
    & 25.4 & 31.9 & 19.9 & \textbf{23.2} & \textbf{22.5}
    & 16.0 & \textbf{17.5} & 33.0 & 3.0
    & 8.8 & 22.8 & 18.1
    & 20.2 & -3.3\% \\

    +\spike 
    & \textbf{27.8} & 29.0 & \textbf{20.0} & 20.4 & {19.0}
    & \textbf{19.2} & {16.7} & {32.0} & 18.3
    & \textbf{9.2} & \textbf{25.2} & \textbf{24.9}
    & \textbf{21.8} & \textbf{+4.3\%} \\

    \midrule
    Qwen
    & 30.9 & 36.2 & 17.7 & 24.6 & 14.9
    & 13.5 & 19.9 & 25.5 & 14.4
    & \textbf{27.8} & \textbf{32.9} & \textbf{32.9}
    & 24.3 & - \\

    +PQ.
    & 31.1 & 42.6 & 21.7 & \textbf{28.4} & \textbf{25.1}
    & 14.5 & \textbf{26.6} & \textbf{27.1} & \textbf{17.4}
    & 8.7 & 22.9 & 24.9
    & 24.3 & +0.0\% \\
    
    +Sum.
    & \textbf{36.6} & \textbf{46.7} & 8.9 & 14.4 & 18.9
    & 5.0 & 24.4 & 22.6 & 0.6
    & 2.5 & 24.8 & 20.8
    & 18.8 & -22.6\% \\

    +\spike 
    & {32.4} & {41.2} & \textbf{23.7} & {25.7} & {24.7}
    & \textbf{16.0} & {23.7} & {26.3} & {16.7}
    & 12.5 & 27.1 & 31.0
    & \textbf{25.1} & \textbf{+3.3\%} \\
    
    \bottomrule
    \end{tabular}
}
\caption{Performance comparison of different document expansion methods on reasoning-intensive retrieval tasks. We compare \spike against pseudo query-based and summary-based document expansion approaches. PQ. denote the pseudo query-based approach and Sum. denote the summary-based approach.}
\label{tab:compareexisting}
\end{table}

\section{More experiment \& analysis result}
\label{sec:moreresult}
\subsection{Reasoning-augmented query result}
\label{subsec:fullanalysis}
To provide a more comprehensive view of how \spike interacts with reasoning-augmented queries, we present the full results of experiments using GPT-4-generated reasoning queries from the BRIGHT benchmark. 
Table~\ref{tab:gpt4query_results} reports nDCG@10 scores across all 12 datasets in BRIGHT, comparing standard retrieval models with and without \spike. 
These results confirm that \spike consistently improves retrieval performance even when applied to queries that already include explicit reasoning. 
It demonstrates not only the robustness of \spike across query types, but also its potential to be effectively integrated into future methods aimed at enhancing reasoning-intensive retrieval performance.
\begin{table}[h!]
\setlength{\tabcolsep}{4pt}
\centering

\resizebox{\textwidth}{!}{
    \begin{tabular}{lcccccccccccccc}
    \toprule
    & \multicolumn{5}{c}{\textbf{Natural language}} 
    & \multicolumn{4}{c}{\textbf{Code}} 
    & \multicolumn{3}{c}{\textbf{Math}} 
    & \multirow{2}{*}{\centering \textbf{{Avg.}}} & \multirow{2}{*}{\centering \textbf{Improv.}} \\
    \cmidrule(r){2-6} \cmidrule(r){7-10} \cmidrule(r){11-13}
    & \textbf{Bio.} & \textbf{Earth.} & \textbf{Econ.} & \textbf{Psy.} & \textbf{Sus.} 
    & \textbf{Rob.} 
    & \textbf{Stack.} & \textbf{Leet.} & \textbf{Pony} 
    & \textbf{Aops} & \textbf{TheoQ.} & \textbf{TheoT.} \\
    \midrule
    E5-Mistral 
    & 29.6 & 43.6 & 20.1 & 26.7 & \textbf{15.6}
    & 11.8 & 17.7 & 29.1 & \textbf{9.0}
    & 5.3 & 25.6 & \textbf{35.7}
    & 22.5 & \multirow{2}{*}{\centering \textbf{{+8.4\%}}} \\
    
    +\spike 
    & \textbf{37.4} & \textbf{46.2} & \textbf{22.3} & \textbf{27.1} & 15.5
    & \textbf{13.4} & \textbf{23.2} & \textbf{30.2} & 8.5
    & \textbf{6.8} & \textbf{28.0} & 34.8
    & \textbf{24.4} & \\
    \cdashline{1-15}
    \addlinespace[2pt]
    
    SFR 
    & 26.2 & 39.1 & 21.5 & 28.3 & \textbf{19.5}
    & 13.4 & 16.8 & 28.4 & 1.5
    & 7.1 & 25.9 & 33.2
    & 21.7 & \multirow{2}{*}{\centering \textbf{{+10.7\%}}}\\
    
    +\spike 
    & \textbf{30.1} & \textbf{40.2} & \textbf{24.7} & \textbf{29.6} & 17.8
    & \textbf{15.2} & \textbf{22.7} & \textbf{30.0} & \textbf{9.4}
    & \textbf{8.8} & \textbf{28.0} & \textbf{34.5}
    & \textbf{24.3} & \\
    \cdashline{1-15}
    \addlinespace[2pt]
    
    GRIT 
    & 33.1 & \textbf{38.9} & 22.3 & 28.8 & 24.1
    & 17.4 & 17.7 & 31.8 & \textbf{11.7}
    & 6.7 & 26.3 & 29.5
    & 24.0 & \multirow{2}{*}{\centering \textbf{{+4.2\%}}}\\
    
    +\spike 
    & \textbf{29.8} & 35.1 & \textbf{23.8} & \textbf{29.1} & \textbf{24.2}
    & \textbf{18.4} & \textbf{22.0} & \textbf{34.2} & 11.4
    & \textbf{8.3} & \textbf{30.0} & \textbf{33.6}
    & \textbf{25.0} & \\
    \cdashline{1-15}
    \addlinespace[2pt]
    
    Qwen
    & 35.8 & \textbf{43.0} & 24.3 & \textbf{34.3} & 24.4
    & 15.6 & 19.7 & 25.4 & 5.2
    & 4.6 & 28.0 & 33.7
    & 24.5 & \multirow{2}{*}{\centering \textbf{{+6.9\%}}}\\
    
    +\spike 
    & \textbf{37.5} & 42.2 & \textbf{26.6} & 33.4 & \textbf{24.5}
    & \textbf{17.6} & \textbf{26.6} & \textbf{28.4} & \textbf{4.3}
    & \textbf{7.3} & \textbf{31.0} & \textbf{35.0}
    & \textbf{26.2} & \\
    
    \bottomrule
    \end{tabular}
}
\caption{The retrieval performance of existing retrieval models and our \spike framework on the BRIGHT benchmark with GPT4 reasoning query provided in \cite{Su2024BRIGHTAR}.}
\label{tab:gpt4query_results}
\end{table}

\subsection{RAG Experimental Result}
\label{subsec:moreRAG}
We provide the full result tables for the RAG experiments conducted in Section~\ref{sec:experiments}. 
Table~\ref{tab:rag_claudegen} presents the complete results using Claude-3.5-sonnet as the generation model and GPT-4o for answer evaluation. 
To further verify the robustness of our findings under different model configurations, we also include results where Llama3.3-70B-Instruct is used as the generation model (Table~\ref{tab:rag_llamagen}). 
This another setup allows us to assess whether the benefits of \spike’s additional context hold consistently across different generation model.

\begin{table}[h!]
\centering
\resizebox{0.99\textwidth}{!}{
\begin{tabular}{lcccccccccccccccc}
\toprule
 \multirow{2}{*}{\centering \textbf{{Ret.}}} & \multicolumn{2}{c}{\textbf{Bio.}} & \multicolumn{2}{c}{\textbf{Earth.}} & \multicolumn{2}{c}{\textbf{Econ.}} & \multicolumn{2}{c}{\textbf{Psy.}} & \multicolumn{2}{c}{\textbf{Rob.}} & \multicolumn{2}{c}{\textbf{Stack.}} & \multicolumn{2}{c}{\textbf{Sus.}}& \multicolumn{2}{c}{\textbf{Avg.}} \\
 & doc. & +sce. & doc. & +sce. & doc. & +sce. & doc. & +sce. & doc. & +sce. & doc. & +sce. & doc. & +sce. & doc. & +sce. \\
\midrule
SBERT & 58.4 & - & 62.4 & - & 52.7 & - & 56.7 & - & 53.3 & - & 66.8 & - & 53.5 & - & 57.7 & - \\
+\spike & 58.8 & \textbf{61.2} & 64.0 & \textbf{64.9} & \textbf{55.7} & 54.4 & 57.0 & \textbf{57.0} & 53.8 & \textbf{55.6} & 65.0 & \textbf{66.1} & 54.7 & \textbf{56.9} &  58.4 & \textbf{59.4} \\
\cdashline{1-17}
\addlinespace[2pt]

SFR & 60.8 & - & 64.1 & - & 54.8 & - & 60.2 & - & 54.1 & - & 67.3 & - & 55.4 & - & 59.5 & - \\
+\spike & 60.5 & \textbf{64.8} &  68.4 & \textbf{70.1} & 54.8 & \textbf{55.6} & \textbf{62.1} & 61.5 & \textbf{56.3} & 55.8 & 67.6 & \textbf{68.6} & 57.1 & \textbf{57.3} & 61.0 & \textbf{62.0} \\
\cdashline{1-17}
\addlinespace[2pt]

Qwen & 62.2 & - & 68.6 & - & 55.9 & - & 60.8 & - & 50.5 & - & 67.7 & - & 56.9 & - & 60.4 & - \\
+\spike & 62.9 & \textbf{65.7} & \textbf{69.4} & 68.7 & 56.7 & \textbf{58.8} & \textbf{60.8} & 60.5 & 55.0 & \textbf{56.1} & 66.9 & \textbf{68.5} & 57.4 & \textbf{58.8} & 61.3 & \textbf{62.5} \\
\midrule
Oracle & \multicolumn{2}{c}{\textit{67.9}} & \multicolumn{2}{c}{\textit{73.3}} & \multicolumn{2}{c}{\textit{66.1}} & \multicolumn{2}{c}{\textit{73.2}} & \multicolumn{2}{c}{\textit{71.0}} & \multicolumn{2}{c}{\textit{77.0}} & \multicolumn{2}{c}{\textit{64.0}} & \multicolumn{2}{c}{\textit{70.3}} \\
\bottomrule
\end{tabular}
}
\caption{Full RAG performance results using Claude-3.5-sonnet as the generation model and GPT-4o for evaluation. doc. denotes document only, which only use retrieved document as context and +sce denotes +Scenario which additionally uses scenario information as context.
}
\label{tab:rag_claudegen}
\end{table}
\begin{table}[h!]
\centering
\resizebox{0.99\textwidth}{!}{
\begin{tabular}{lcccccccccccccccc}
\toprule
 \multirow{2}{*}{\centering \textbf{{Ret.}}} & \multicolumn{2}{c}{\textbf{Bio.}} & \multicolumn{2}{c}{\textbf{Earth.}} & \multicolumn{2}{c}{\textbf{Econ.}} & \multicolumn{2}{c}{\textbf{Psy.}} & \multicolumn{2}{c}{\textbf{Rob.}} & \multicolumn{2}{c}{\textbf{Stack.}} & \multicolumn{2}{c}{\textbf{Sus.}}& \multicolumn{2}{c}{\textbf{Avg.}} \\
 & doc. & +sce. & doc. & +sce. & doc. & +sce. & doc. & +sce. & doc. & +sce. & doc. & +sce. & doc. & +sce. & doc. & +sce. \\
\midrule
SBERT & 56.5 & - & 57.8 & - & 53.2 & - & 55.0 & - & 53.0 & - & 64.1 & - & 51.9 & - & 55.9 & - \\
+\spike & 59.1 & \textbf{61.0} & 63.2 & \textbf{65.1} & \textbf{56.3} & 55.0 & \textbf{57.6} & 56.9 & 52.3 & \textbf{57.1} & 64.9 & \textbf{66.5} & 55.4 & \textbf{57.4} &  58.4 & \textbf{59.9} \\
\cdashline{1-17}
\addlinespace[2pt]

SFR & 57.6 & - & 60.2 & - & 53.7 & - & 59.1 & - & 56.4 & - & 69.3 & - & 49.4 & - & 57.9 & - \\
+\spike & 60.8 & \textbf{65.0} &  66.7 & \textbf{69.6} & 55.3 & \textbf{57.2} & \textbf{62.2} & 62.0 & \textbf{56.1} & 55.8 & 67.6 & \textbf{67.8} & 55.6 & \textbf{57.5} & 60.6 & \textbf{62.1} \\
\cdashline{1-17}
\addlinespace[2pt]

Qwen & 60.6 & - & 62.4 & - & 51.7 & - & 57.5 & - & 51.7 & - & 65.2 & - & 53.3 & - & 57.5 & - \\
+\spike & 63.2 & \textbf{65.8} & 68.7 & 69.4 & 57.1 & \textbf{58.2} & \textbf{61.8} & 61.2 & 52.5 & \textbf{58.4} & 66.2 & \textbf{68.7} & 58.1 & \textbf{58.3} & 61.1 & \textbf{62.9} \\
\midrule
Oracle & \multicolumn{2}{c}{\textit{66.4}} & \multicolumn{2}{c}{\textit{73.8}} & \multicolumn{2}{c}{\textit{65.0}} & \multicolumn{2}{c}{\textit{73.0}} & \multicolumn{2}{c}{\textit{72.6}} & \multicolumn{2}{c}{\textit{76.3}} & \multicolumn{2}{c}{\textit{63.7}} & \multicolumn{2}{c}{\textit{70.1}} \\
\bottomrule
\end{tabular}
}
\caption{Full RAG performance results using Llama3.3-70B-Instruct as the generation model and GPT-4o for evaluation. doc. denotes document only, which only use retrieved document as context and +sce denotes +Scenario which additionally uses scenario information as context.
}
\label{tab:rag_llamagen}
\end{table}

\subsection{Analysis result}
\label{subsec:fullanalysis}
Table~\ref{tab:analysis_full} presents the full results of our analysis, as discussed in Section~\ref{sec:analysis}.

\section{Experiment details}
\label{sec:appendix}
\begin{table}[t!]
\setlength{\tabcolsep}{4pt}
\centering

\resizebox{\textwidth}{!}{
    \begin{tabular}{lc|ccccccccccccc}
    \toprule
    \multirow{2}{*}{\centering \textbf{Comp.}}
    & \multirow{2}{*}{\centering \textbf{$\alpha$}}
    & \multicolumn{5}{c}{\textbf{Natural language}} 
    & \multicolumn{4}{c}{\textbf{Code}} 
    & \multicolumn{3}{c}{\textbf{Math}} 
    & \multirow{2}{*}{\centering \textbf{{Avg.}}} \\
    \cmidrule(r){3-7} \cmidrule(r){8-11} \cmidrule(r){12-14}
    &
    & \textbf{Bio.} & \textbf{Earth.} & \textbf{Econ.} & \textbf{Psy.} & \textbf{Sus.} 
    & \textbf{Rob.} 
    & \textbf{Stack.} & \textbf{Leet.} & \textbf{Pony} 
    & \textbf{Aops} & \textbf{TheoQ.} & \textbf{TheoT.} \\
    \midrule
    \multirow{9}{*}{\centering M}
    & 0.0
    & 18.9 & 20.6 & 13.7 & 18.2 & 12.2 & 6.5 & 11.8 & 17.1 & 2.7 & 1.4 & 11.1 & 18.7 & 12.7 \\

    & 0.1
    & 21.0 & 22.7 & 14.4 & 19.7 & 13.3 & 7.7 & 11.7 & 18.9 & 3.6 & 1.6 & 13.4 & 19.3 & 13.9 \\

    & 0.2
    & 22.4 & 24.5 & 15.0 & 20.5 & 14.5 & 8.7 & 12.3 & 21.7 & 4.0 & 2.0 & 15.3 & 20.6 & 15.1 \\

    & 0.3
    & 23.1 & 26.7 & 16.1 & 21.0 & 17.0 & 9.6 & 12.7 & 23.9 & 5.2 & 2.2 & 16.5 & 23.0 & 16.4 \\
    
    & 0.4
    & 23.9 & 27.9 & 17.6 & 21.9 & 17.9 & 9.9 & 13.1 & 25.3 & 7.1 & 2.7 & 18.5 & 23.3 & 17.4 \\

    & 0.5
    & 24.6 & 28.8 & 18.2 & 22.5 & 18.6 
    & 10.9 & 13.0 & 27.1 & 9.2 
    & 3.4 & 20.0 & 24.7 
    & 18.4 \\

    & 0.6
    & 25.2 & 30.0 & 18.7 & 22.0 & 19.7 
    & 12.5 & 14.0 & 28.8 & 10.8 
    & 4.5 & 21.5 & 27.1 
    & 19.5 \\

    & 0.7
    & 24.8 & 29.3 & 19.5 & 21.4 & 19.4 
    & 14.0 & 14.1 & 29.3 & 12.5 
    & 5.5 & 22.7 & 17.3 
    & 20.0 \\

    & 0.8
    & 24.7 & 28.2 & 18.0 & 19.6 & 20.1 
    & 15.1 & 13.9 & 29.1 & 13.6 
    & 6.3 & 24.1 & 27.1
    & 20.0 \\

    & 0.9
    & 23.6 & 28.6 & 18.0 & 18.2 & 19.5 
    & 16.4 & 13.6 & 29.2 & 16.7 
    & 6.7 & 25.8 & 27.0
    & 20.3 \\

    \midrule
    \multirow{9}{*}{\centering I}
    & 0.0
    & 22.9 & 22.6 & 15.1 & 17.2 & 12.5 & 10.2 & 14.6 & 19.6 & 7.3 & 1.3 & 12.0 & 13.7 & 14.1 \\

    & 0.1
    & 23.9 & 24.4 & 16.5 & 18.9 & 14.2 & 11.1 & 14.8 & 22.3 & 8.2 & 2.0 & 13.9 & 15.6 & 15.5 \\

    & 0.2
    & 25.2 & 26.3 & 17.7 & 20.1 & 15.2 & 12.2 & 14.7 & 24.9 & 10.0 & 2.5 & 16.0 & 16.2 & 16.7 \\

    & 0.3
    & 26.1 & 27.7 & 17.5 & 20.6 & 16.0 & 12.6 & 14.3 & 26.2 & 12.0 & 3.4 & 18.3 & 16.9 & 17.6 \\
    
    & 0.4
    & 26.8 & 28.8 & 18.0 & 20.3 & 16.7 & 12.9 & 13.7 & 27.9 & 14.4 & 4.0 & 19.9 & 18.2 & 18.4 \\

    & 0.5
    & 26.5 & 30.0 & 17.9 & 20.7 & 17.3
    & 14.7 & 14.7 & 28.7 &  15.9
    & 4.9 & 20.8 & 19.1
    & 19.3 \\

    & 0.6
    & 25.9 & 31.1 & 18.5 & 20.1 & 17.6
    & 15.5 & 15.5 & 29.3 & 16.7
    & 5.4 & 21.8 & 22.7
    & 20.0 \\

    & 0.7
    & 24.9 & 30.6 & 18.3 & 19.3 & 18.3
    & 16.7 & 15.7 & 29.4 & 16.5
    & 6.0 & 22.3 & 25.2
    & 20.3 \\

    & 0.8
    & 18.8 & 18.7 & 29.1 & 18.7 & 18.6
    & 16.5 & 14.4 & 29.3 & 15.7 
    & 6.7 & 23.7 & 25.1
    & 20.2 \\

    & 0.9
    & 23.5 & 28.6 & 18.1 & 18.1 &  18.7
    & 17.1 & 14.0 & 28.7 &  14.7
    & 7.2 & 25.1 & 26.1
    & 20.0 \\

    \midrule
    \multirow{9}{*}{\centering E}
    & 0.0
    & 27.2 & 27.6 & 16.2 & 20.7 & 14.0 & 8.3 & 12.6 & 20.2 & 15.9 & 0.4 & 10.9 & 14.0 & 15.7 \\

    & 0.1
    & 27.7 & 28.6 & 16.7 & 21.7 & 14.5 & 9.2 & 13.0 & 22.6 & 16.3 & 0.6 & 13.0 & 14.8 & 16.6 \\

    & 0.2
    & 28.0 & 30.1 & 16.9 & 22.3 & 15.2 & 10.1 & 13.5 & 25.2 & 16.5 & 1.1 & 15.5 & 18.8 & 17.8 \\

    & 0.3
    & 28.7 & 30.9 & 17.4 & 23.8 & 16.8 & 11.3 & 14.3 & 26.7 & 16.2 & 1.4 & 17.2 & 23.2 & 19.0 \\
    
    & 0.4
    & 29.2 & 32.0 & 17.7 & 24.4 & 18.3 & 12.7 & 14.4 & 27.7 & 17.3 & 2.4 & 18.5 & 24.9 & 20.0 \\
    
    & 0.5
    & 29.4 & 33.0 & 17.8 & 24.0 & 19.2
    & 13.6 & 15.1 & 28.8 & 17.3
    & 3.5 & 20.6 & 27.9
    & 20.9 \\

    & 0.6
    & 29.0 & 32.6 & 19.0 & 22.6 & 19.3
    & 15.0 & 15.3 & 30.4 & 18.2
    & 4.7 & 21.9 & 28.2
    & 21.3 \\

    & 0.7
    & 28.3 & 32.3 & 19.2 & 22.2 & 19.7
    & 16.1 & 15.6 & 30.4 & 18.0
    & 5.1 & 23.8 & 28.5
    & 21.6 \\

    & 0.8
    & 26.5 & 32.0 & 18.5 & 20.3 & 19.6
    & 16.8 & 15.0 & 29.8 & 17.0
    & 6.5 & 24.8 & 28.1
    & 21.2 \\

    & 0.9
    & 25.0 & 30.1 & 17.9 & 19.4 & 19.7
    & 16.6 & 14.0 & 29.3 & 16.6 
    & 7.1 & 25.5 & 26.5
    & 20.6 \\

    \midrule
    \multirow{9}{*}{\centering M+E}
    & 0.0
    & 25.9 & 29.9 & 15.7 & 21.3 & 17.9 & 15.9 & 15.8 & 23.4 & 17.9 & 1.2 & 14.5 & 18.5 & 18.2 \\

    & 0.1
    & 26.4 & 30.2 & 16.2 & 22.1 & 19.1 & 16.6 & 16.5 & 24.8 & 18.0 & 1.9 & 16.2 & 19.9 & 19.0 \\

    & 0.2
    & 27.4 & 31.5 & 16.4 & 23.0 & 19.6 & 17.1 & 16.7 & 26.7 & 18.0 & 2.4 & 17.7 & 21.2 & 19.8 \\

    & 0.3
    & 28.1 & 32.2 & 16.8 & 23.7 & 20.3 & 17.8 & 16.0 & 28.4 & 18.1 & 3.3 & 18.5 & 23.3 & 20.5 \\
    
    & 0.4
    & 28.0 & 34.0 & 17.0 & 23.9 & 19.9 & 18.3 & 16.7 & 29.4 & 18.1 & 4.3 & 20.7 & 23.7 & 21.2 \\
    
    & 0.5
    & 27.9 & 34.4 & 17.5 & 23.5 & 20.4
    & 18.6 & 16.8 & 29.2 & 18.6
    & 6.5 & 25.8 & 25.8
    & 22.1 \\

    & 0.6
    & 26.6 & 32.9 & 17.7 & 21.4 & 21.0
    & 18.9 & 16.9 & 29.1 & 18.3
    & 6.2 & 22.3 & 28.9
    & 21.7 \\

    & 0.7
    & 25.9 & 33.0 & 18.2 & 20.6 & 20.6
    & 18.4 & 16.2 & 29.4 & 17.5
    & 7.0 & 23.4 & 28.4
    & 21.6 \\

    & 0.8
    & 25.0 & 32.0 & 18.2 & 20.1 & 20.4
    & 17.9 & 14.9 & 29.5 & 15.9
    & 7.5 & 24.5 & 27.1
    & 21.1 \\

    & 0.9
    & 23.9 & 28.8 & 18.1 & 18.7 & 19.6
    & 16.9 & 14.1 & 28.6 & 15.1
    & 7.5 & 25.2 & 26.6
    & 20.3 \\

    \midrule
    \multirow{9}{*}{\centering I+E}
    & 0.0
    & 24.0 & 25.9 & 14.5 & 21.9 & 15.3 & 13.2 & 14.7 & 22.5 & 9.9 & 1.2 & 11.7 & 13.0 & 15.6 \\

    & 0.1
    & 24.3 & 27.8 & 14.8 & 22.4 & 16.0 & 13.4 & 15.0 & 25.0 & 11.7 & 1.4 & 13.4 & 14.8 & 16.7 \\

    & 0.2
    & 25.4 & 29.4 & 15.9 & 22.1 & 18.0 & 14.7 & 14.8 & 26.7 & 13.1 & 1.4 & 15.0 & 16.8 & 17.8 \\

    & 0.3
    & 26.2 & 31.0 & 16.5 & 22.7 & 18.5 & 15.9 & 15.0 & 27.7 & 14.0 & 2.2 & 16.6 & 19.3 & 18.8 \\
    
    & 0.4
    & 26.9 & 32.0 & 16.9 & 23.0 & 19.2 & 16.8 & 16.1 & 27.7 & 14.7 & 3.0 & 18.8 & 21.1 & 19.7 \\
    
    & 0.5
    & 27.1 & 32.3 & 17.6 & 21.9 & 19.5
    & 16.7 & 16.1 & 28.7 & 15.3
    & 4.3 & 20.3 & 24.0
    & 20.3 \\

    & 0.6
    & 27.0 & 32.5 & 18.1 & 21.7 & 19.6
    & 16.9 & 16.0 & 29.4 & 15.7
    & 4.9 & 21.6 & 25.5
    & 20.7 \\

    & 0.7
    & 25.7 & 31.8 & 18.2 & 20.9 & 19.6
    & 17.5 & 16.0 & 29.7 & 15.5
    & 5.7 & 22.8 & 26.3
    & 20.8 \\

    & 0.8
    & 24.8 & 31.0 & 17.9 & 20.2 & 19.2
    & 17.5 & 14.8 & 30.3 &  14.9
    & 7.0 & 24.0 & 27.3
    & 20.7 \\

    & 0.9
    & 23.2 & 29.2 & 17.8 & 18.5 & 19.0 
    & 16.7 & 14.2 & 29.0 & 14.2
    & 6.9 & 25.0 & 26.6
    & 20.0 \\
    \bottomrule
    \end{tabular}
}
\caption{Full ablation results presented in Section~\ref{sec:analysis}.}
\label{tab:analysis_full}
\end{table}

\subsection{Dataset}
\label{appendix:evaldataset}
\subsubsection{BRIGHT}
BRIGHT includes 1,398 real-world queries covering diverse domains such as economics, psychology, robotics, mathematics, and software programming. 
These queries are carefully designed to reflect challenging scenarios that demand deep comprehension and reasoning to retrieve relevant documents.
\cite{Su2024BRIGHTAR} categorizes datasets into groups such as StackExchange, Coding, and Theorem-based collections. 
Specifically, individual datasets are classified as follows:
\begin{itemize}[leftmargin=*,topsep=2pt,itemsep=2pt,parsep=0pt]
  \item StackExchange: Biology (\textbf{Bio.}), Earth Science (\textbf{Earth.}), Economics (\textbf{Econ.}), Psychology (\textbf{Psy.}), Robotics (\textbf{Rob.}), Stack Overflow (\textbf{Stack.}), Sustainable Living (\textbf{Sus.}),
  \item Coding: Leetcode (\textbf{Leet.}), Pony (\textbf{Pony})
  \item Theorem-based: Aops (\textbf{AoPS}), TheoremQA-Question (\textbf{TheoQ.}), TheoremQA-Theorem (\textbf{TheoT.})
\end{itemize}
Our work adopts a different classification based on document type, categorizing datasets into Natural Language, Code, and Math to better capture the retrieval challenges associated with different content structures.
Specifically, individual datasets are classified as follows:
\begin{itemize}[leftmargin=*,topsep=2pt,itemsep=2pt,parsep=0pt]
  \item Natural Language: Biology (\textbf{Bio.}), Earth Science (\textbf{Earth.}), Economics (\textbf{Econ.}), Psychology (\textbf{Psy.}), Sustainable Living (\textbf{Sus.})
  \item Code: Leetcode (\textbf{Leet.}), Pony (\textbf{Pony}), Robotics (\textbf{Rob.}), Stack Overflow (\textbf{Stack.}),
  \item Math: Aops (\textbf{AoPS}), TheoremQA-Question (\textbf{TheoQ.}), TheoremQA-Theorem (\textbf{TheoT.})
\end{itemize}
During the evaluation process, for instruction-following models used in our experiments, we directly utilized the instructions provided by \cite{Su2024BRIGHTAR}.

\subsubsection{BEIR}
BEIR is a benchmark comprising diverse information retrieval tasks, consisting of 18 datasets across various domains such as Wikipedia, scientific publications, and others.
In this work, we use the corpus from 15 publicly available BEIR dataset (MS MARCO, TREC-COVID, NFCorpus, NQ, HotpotQA, FiQA-2018, ArguAna, Touche-2020, CQADupStack, Quora, DBPedia, SCIDOCS, FEVER, Climate-FEVER, and SciFact) to train our scenario generator.

\subsection{Implementation details}
\label{appendix:implementationdetails}

\subsubsection{Scenario-augmented training dataset}
\label{appendix:Scenarioaugmentedtrainingset}
Scenario generation for constructing the scenario-augmented training dataset was performed using GPT-4o with greedy decoding to ensure consistent and high-quality outputs. 
Additionally, to guarantee that each scenario component was generated without omission, we leveraged OpenAI's structured output feature, ensuring that all output scenarios should be JSON format.
For the prompt used in this process, please refer to Table~\ref{tab:dataset_prompt} in \ref{appendix:prompt}.

For each individual document, we did not set a fixed number of scenarios to be generated. 
Instead, we employ an adaptive approach, where the number of generated scenarios is determined based on the content of each document. 
Documents with more extensive content resulted in a larger number of generated scenarios, while those with less content generated fewer scenarios.

\subsubsection{Scenario generator}
\label{appendix:Scenariogenerator}
The scenario generator is optimized using AdamW with a learning rate of 2e-5, a linear warmup scheduler, weight decay of 0.1, and a batch size of 4 with gradient accumulation of 2. 
Optimization is conducted for a maximum of 10 epochs, with early stopping based on evaluation loss and a patience of 4.
The LoRA hyperparameters are set as follows: r = 32, alpha=64, dropout = 0.1. 
After training the scenario generator, we applied it to the entire BRIGHT corpus to generate scenarios for each dataset. 

\subsection{Reasoning-augmented query}
\label{appendix:reasoningaugmentedquery}
BRIGHT provides reasoning-augmented queries, which are reformulated versions of the original queries generated using various large language models.
These queries incorporate explicit step-by-step reasoning, aiming to clarify the user intent and better guide retrieval models.
Among the available variants, we adopt the GPT-4–generated reasoning queries, which achieved the best performance in the experiments reported in the \cite{Su2024BRIGHTAR}.

\subsection{Criteria for human evaluation}
\label{appendix:criteriahumaneval}
\begin{figure}[h]
    \centering
    \includegraphics[width=\columnwidth]{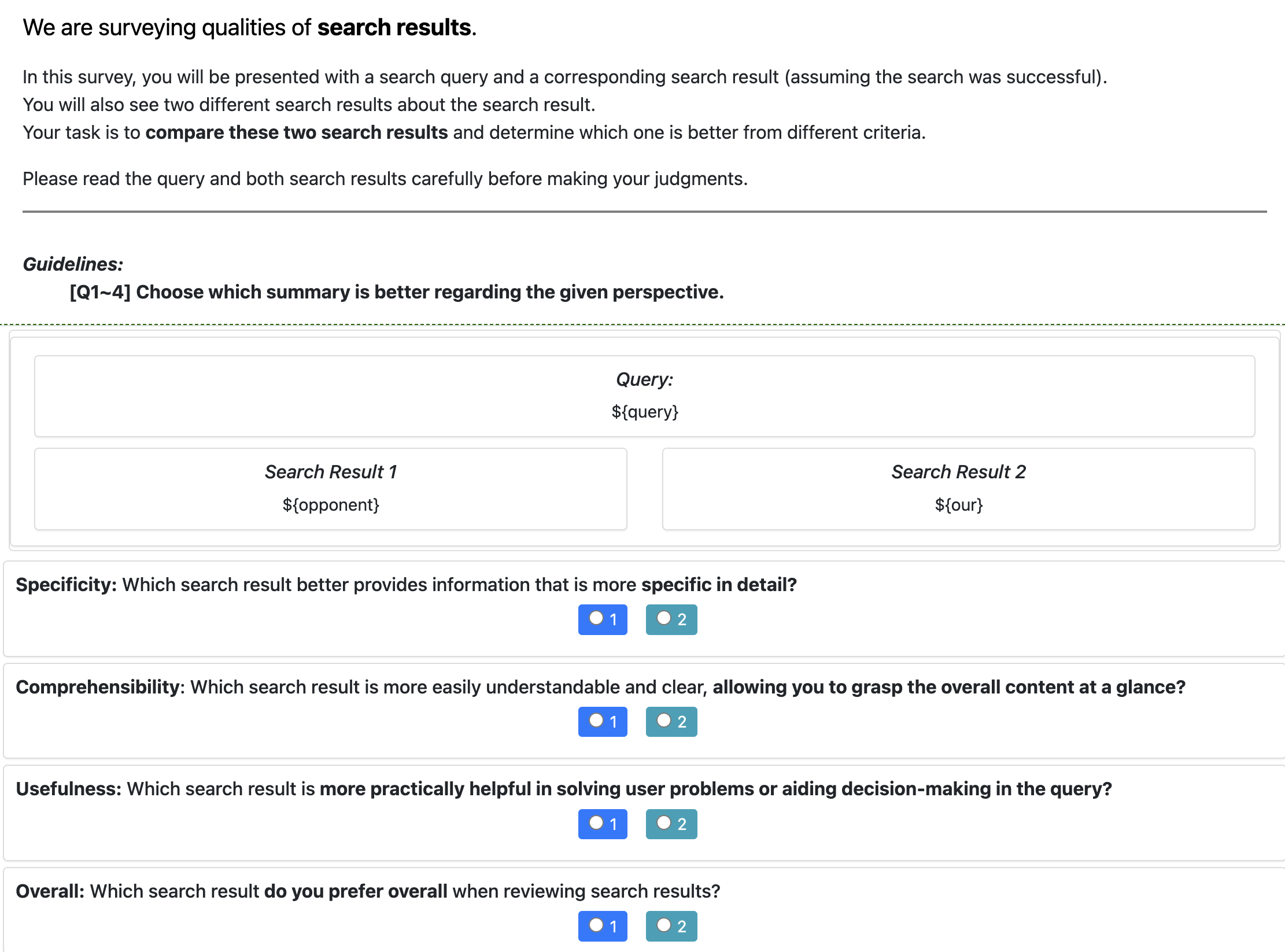}
    \caption{The interface for human evaluation}
    \label{fig:humanevalinterface}
\end{figure}

We randomly sample 100 examples from the BRIGHT test set and ask three human judges per example to compare different retrieval contexts following four criteria: 
\begin{itemize}[leftmargin=*,topsep=2pt,itemsep=2pt,parsep=0pt]
  \item \textbf{Specificity}: Which search result better provides information that is more specific in detail?
  \item \textbf{Comprehensibility}: Which search result is more easily understandable and clear, allowing you to grasp the overall content at a glance?
  \item \textbf{Usefulness}: Which search result is more practically helpful in solving user problems or aiding decision-making in the query?
  \item \textbf{Overall}: Which search result do you prefer overall when reviewing search results?
\end{itemize}
We show the interface for the evaluation in Figure~\ref{fig:humanevalinterface}.

\subsection{Experimental setting for RAG experiment}
\label{appendix:RAGexperimentsetting}
In Section~\ref{subsec:effectiveRAG}, we investigate whether \spike can serve as an effective context provider in retrieval-augmented generation (RAG) by offering additional information alongside retrieved documents. 
In this experiment, we use the queries and reference answers from the BRIGHT dataset, following the experimental setup proposed by \citet{Su2024BRIGHTAR}, and use GPT-4o as the evaluation model to score the generated answers.
Specifically, in the oracle setting, the full set of gold documents corresponding to each query is provided as context, while the retrieval setting uses the top-10 documents retrieved by the retrieval model. 
The prompts used for question answering and evaluation also follow those introduced in \cite{Su2024BRIGHTAR}.

\subsection{Prompt}
\label{appendix:prompt}
We present four types of prompts used in our experiments:
\begin{itemize}[leftmargin=*,topsep=2pt,itemsep=2pt,parsep=0pt]
  \item \textbf{Construct scenario-augmented dataset}: The prompt designed for constructing scenario-augmented dataset is shown in Table~\ref{tab:dataset_prompt}
  \item \textbf{Scenario generator instruction}: The prompt designed for training scenario generator is shown in Table~\ref{tab:generator_prompt}
  \item \textbf{RAG answering}: The prompt designed for answering in RAG setting is shown in Table~\ref{tab:rag_answering_prompt}
  \item \textbf{Evaluate RAG}: The prompt designed for evaluating answer in RAG setting is shown in Table~\ref{tab:rag_evaluation_prompt}
\end{itemize}

\newpage

\begin{table}[h]
    \small
    \centering
    \begin{tabular}{p{14cm}}
    \toprule
    \textbf{Prompt for constructing scenario-augmented dataset} \\
    \midrule
\textcolor{teal}{\textbf{[Task Description]}}\\
You are an advanced language model specializing in knowledge extraction and user need modeling. Your task is to extract hypothetical user scenarios from a given {dataset} document, ensuring that the generated information needs reflect the document's overall insights and knowledge, rather than isolated details.
\\\\
**Step 1: Document Analysis**: \\
Summarize the key points of the document in a structured manner. This step should not be a direct extraction but should synthesize the document's core concepts, key arguments, and insights, avoiding specific code snippets, variable names, or minor details. 
\\
Content: \\
- Main Topic: Briefly describe the primary subject of the document \\
- Key Aspects: Summarize the core concepts, insights, or knowledge presented \\
\\
**Step 2: Generate Possible Information Needs**: \\
Based on the document analysis, generate a diverse set of possible information needs that can be satisfied by the document, ensuring that they **focus on high-level insights, generalizable knowledge, or core principles conveyed by the document rather than specific implementation details (e.g., function names, variable names, or isolated sections)**.
\\
Guidelines: \\
- The information needs must align with the document's main message and core knowledge, not minor details. \\
- Focus on concepts, reasoning, and insights rather than localized facts.\\
- Ensure that they **focus on high-level insights, generalizable knowledge, or core principles conveyed by the document rather than specific implementation details (e.g., function names, variable names, or isolated sections)**.\\
- Ensure that the needs **capture different aspects of the document's knowledge** rather than concentrating on a single part.\\
\\
Format:\\
- Each information need is started with "A User wants to know"\\
- Generate a python list of information needs. (e.g. ["information need 1", "information need 2", "information need 3"])\\

\\
**Step 3: Generate Explanation for Each Information Need**:\\
For each information need, explain how the document fulfills that need, ensuring that explanations are generalized and conceptual rather than overly detailed. Avoid focusing on function names, variable names, or specific lines unless absolutely necessary for clarity.\\
\\
Format:\\
- Generate JSON format with the following components: \\
- Key: information need \\
- Value: explanation for the information need \\

\\\\
\textcolor{teal}{\textbf{[Text Content]}}\\
...
\\
\bottomrule
    \end{tabular}
    \caption{The prompt for constructing scenario-augmented dataset}
    \label{tab:dataset_prompt}
\end{table}

\begin{table}[h]
    \small
    \centering
    \begin{tabular}{p{14cm}}
    \toprule
    \textbf{Scenario generator instruction} \\
    \midrule
\textcolor{teal}{\textbf{[Task Description]}}\\
You are an advanced language model specializing in knowledge extraction and user need modeling. Your task is to extract hypothetical user scenarios from a given {dataset} document, ensuring that the generated information needs reflect the document's overall insights and knowledge, rather than isolated details.
\\\\
Content:\\
- Main Topic: Briefly describe the primary subject of the document\\
- Key Aspects: Summarize the core concepts, insights, or knowledge presented\\
- Information Needs: Generate a diverse set of possible information needs that can be satisfied by the document\\
- Explanation: Explain how the document fulfills that need, ensuring that explanations are generalized and conceptual rather than overly detailed.\\
\\
Format:\\
- Generate JSON format\\

\\\\
\textcolor{teal}{\textbf{[Text Content]}}\\
...
\\
\bottomrule
    \end{tabular}
    \caption{Scenario generator instruction}
    \label{tab:generator_prompt}
\end{table}

\begin{table}[h]
    \small
    \centering
    \begin{tabular}{p{14cm}}
    \toprule
    \textbf{RAG answering Prompt} \\
    \midrule
\textcolor{teal}{\textbf{[Task Description]}}\\
Problem:\\
{question}
\\\\
Document:\\
{document}
\\\\
Based on the provided documents, write an answer to the problem.
\\
\bottomrule
    \end{tabular}
    \caption{The prompt for RAG answering}
    \label{tab:rag_answering_prompt}
\end{table}

\begin{table}[h]
    \small
    \centering
    \begin{tabular}{p{14cm}}
    \toprule
    \textbf{RAG evaluation Prompt} \\
    \midrule
\textcolor{teal}{\textbf{[Task Description]}}\\
———- PROBLEM START ———-\\
{problem}\\
———- PROBLEM END ———-\\
———- STUDENT ANSWER START ———-\\
{predicted answer}\\
———- STUDENT ANSWER END ———-\\
———- REFERENCE ANSWER START ———-\\
{gold answer}\\
———- REFERENCE ANSWER END ———-\\
Criteria:\\
0 - The student's answer is completely irrelevant or blank.\\
10 - The student's answer addresses about 10\% of the reference content.\\
20 - The student's answer addresses about 20\% of the reference content.\\
30 - The student's answer addresses about 30\% of the reference content.\\
40 - The student's answer addresses about 40\% of the reference content.\\
50 - The student's answer addresses about 50\% of the reference content.\\
60 - The student's answer addresses about 60\% of the reference content.\\
70 - The student's answer addresses about 70\% of the reference content.\\
80 - The student's answer addresses about 80\% of the reference content.\\
90 - The student's answer addresses about 90\% of the reference content.\\
100 - The student's answer addresses about 100\% of the reference content.\\
Use the following format to give a score:\\
REASON:\\
Describe why you give a specific score\\
SCORE:\\
The score you give, e.g., 60\\
Do not say anything after the score\\
\bottomrule
    \end{tabular}
    \caption{The prompt for RAG evaluation}
    \label{tab:rag_evaluation_prompt}
\end{table}

\end{document}